%% file: main.tex
\documentclass[a4paper,fleqn]{cas-dc}

\usepackage[authoryear,longnamesfirst]{natbib}

\shortcites{%
deep,
sardo,
remote-surgery,
ASCHENBRENNER2015159,
5g-agv}

\usepackage{tikz}
\usetikzlibrary{decorations.pathreplacing}
\usetikzlibrary{shapes.geometric}
\usetikzlibrary{calc}
\usetikzlibrary{arrows}
\usetikzlibrary{decorations.markings}
\usetikzlibrary{shapes.geometric}
\usetikzlibrary{patterns,snakes}
\usetikzlibrary{decorations.text}
\usetikzlibrary{arrows.meta}
\usepackage{pgfplots}
\pgfplotsset{compat=newest}
\usepgfplotslibrary{fillbetween}

\usepackage{amsmath,amssymb,amsfonts,amsthm}
\usepackage{mathtools}

\usepackage{subfig}
\usepackage{standalone}
\usepackage[linesnumbered,ruled,vlined]{algorithm2e}
\usepackage[font=small]{caption}

\usepackage{listings}
\DeclareMathOperator*{\argmin}{arg\,min}

\newtheorem{theorem}{Theorem}
\newtheorem{lemma}[theorem]{Lemma}

\def\1{\mathbf{1}}

\newcounter{nalg} 
\renewcommand{\thenalg}{\arabic{nalg}} 
\DeclareCaptionLabelFormat{algocaption}{Algorithm \thenalg} 

\def\tsc#1{\csdef{#1}{\textsc{\lowercase{#1}}\xspace}}
\tsc{WGM}
\tsc{QE}

\begin{document}
\let\WriteBookmarks\relax
\def\floatpagepagefraction{1}
\def\textpagefraction{.001}

\shorttitle{Modelling M/M/R-JSQ-PS sojourn time distribution for URLLC services}    

\shortauthors{Geraint I. Palmer, Jorge Martín-Pérez}  

\title [mode = title]{Modelling M/M/R-JSQ-PS sojourn time distribution for\\Ultra-Reliable Low Latency Communication services}  



%


%
\author[1]{Geraint I. Palmer}[type=editor,
  bioid=1,
  orcid=0000-0001-7865-6964,
  ]

\cormark[1]


\ead{palmergi1@cardiff.ac.uk}





\address[1]{School of Mathematics, Cardiff University,
            Senghennydd Road,
            Cardiff,
            CF24 4AG,
            Wales}

\author[2]{Jorge Martín-Pérez}[type=editor,
  bioid=2,
  orcid=0000-0001-9295-1601
  ]


\ead{jmartinp@it.uc3m.es}




\address[2]{{Department of Telematics Engineering, Universidad Carlos III de Madrid},
            {Av. de la Universidad}, 
            {Leganés}, 
            {Spain}}




\begin{abstract}
    The future Internet promises to support time-sensitive
    services that require ultra low
    latencies and reliabilities of 99.99\%.
    Recent advances in cellular and WiFi connections
    enhance the network to meet high reliability and
    ultra low latencies. However, the aforementioned services
    require that the server processing time 
    ensures low latencies with high reliability, otherwise the
    end-to-end performance is not met.
    To that end, in this paper we use queuing theory to model the
    sojourn time distribution for Ultra-Reliable Low Latency
    Communication services of \mbox{M/M/R-JSQ-PS}
    systems: Markovian queues with $R$ CPU servers following a join shortest
    queue processor-sharing discipline (for example Linux systems). We
    develop open-source simulation software, and develop and compare
    six analytical approximations for the sojourn time distribution.
    The proposed approximations yield Wasserstein distances below
    2~time units, and upon medium loads incur into errors of less than
    1.78~time units (e.g., milliseconds) for the
    99.99\textsuperscript{th} percentile sojourn time.
    Moreover, the proposed sojourn time approximations are stable regardless the
    number of CPUs and stay close to the simulations.

\end{abstract}



\begin{keywords}
queueing \sep simulation
\end{keywords}

\maketitle

\section{Introduction}

Recent advances in the networking
community aim at a better control over
infrastructure behaviour. Although the Internet
was designed to provide a best-effort
delivery~\citep{rfc3724},
5G~\citep{21.915},
WiFi~6~\citep{IEEE802.11ax,IEEE802.11be}
and WiFi~7~\citep{IEEE802.11be}
have enhanced the mobile connectivity to increase
the network reliability. With the
new wireless technologies it is possible
to support services that require low
latencies and high reliabilities like
vehicle to everything
(V2X)~\citep{5g-americas},
drones~control~\citep{sardo},
remote surgery~\citep{remote-surgery},
or Industry~4.0~\citep{ASCHENBRENNER2015159}.
In particular, 
it is now possible to remotely
control an industrial robotic arm over a
wireless connection~\citep{deep} while
guaranteeing a reliable communication.

The 3\textsuperscript{rd} Generation
Partnership Project (3GPP) claims \citep{38.913}
that the aforementioned services require an
Ultra-Reliable Low Latency Communication
(URLLC) over the Internet.
That is, any URLLC service should
foresee Internet latencies in the order of 10~ms
and reliabilities above a 99\%.
5G and WiFi already guarantee a low latency and
reliable wireless communication
through diverse mechanisms
\citep{28.811,21.915,23.725,23.502,29.517},
but it is out of their scope whether
the processing time of Internet traffic
satisfies an URLLC.

Internet packets exchanged by
URLLC services are typically processed in a remote
server accessible through a 5G or WiFi connection.
In a remotely controlled
industrial robot \citep{deep} the steps
are as follows: ($i$) the robot reports
its position within packets sent over
5G/WiFi to a server;
($ii$) the server calculates
the next robot position; and
($iii$) the robot receives an instruction
with its new
position over the 5G/WiFi connection.
Calculating the next robot
position at step ($ii$) induces a
processing latency that depends
on factors such as the server load,
the arrival distribution of sensor data,
how fast the server is,
how many CPUs the server has, or
how complex operations are.

In the case of a remotely controlled robotic
arm, the server CPUs perform
inverse/forward kinematics~\citep{kinematics}
and PID control~\citep{pid} operations to derive
the next position of the robotic arm.
Both operations are performed at a CPU
for each URLLC packet, and their delay
is impacted by the number of 
packets being processed at the same CPU.
Hence, if a CPU is attending
multiple URLLC packets it is more unlikely
that the processing time remains below
the latency requirement of
10~ms. Consequently, the latency
of a remotely controlled robotic arm
may exceed the 10~ms requirement
even if the 5G or WiFi~6/7 connection
provides URLLC -- steps ($i$) and ($iii$)
in the prior paragraph.

Assuming that a 5G or WiFi~6/7 connection
suffices to provide URLLC for a service
is not enough. 
It is also necessary to understand how
the processing time is distributed when
URLLC traffic is attended by a server.
Only when both the URLLC traffic
processing and wireless communication
satisfy the latency requirements, we can tell
that the network infrastructure provides
an URLLC service, e.g., that it ensures latencies
below 10~ms 99\% of probability. Therefore, it is
of paramount importance to model the URLLC processing
latency.

\bigskip
In this paper we study how the traffic processing
latency is distributed to determine whether
a service meets URLLC.
Specifically, we develop open-source simulation
software, and also propose analytical approximations
to characterise the processing latency of
servers that dispatch the traffic processing
to the least loaded CPU within a pool of
$R$ CPUs.
As assumed by the state of the art~\citep{RCohen15,jemaa2016qos,oljira2017model};
and alike Linux-based systems, we assume
that each CPU utilises a processor-sharing
policy to attend the traffic processing.

The contributions of our work are summarised as
follows:
\begin{itemize}
    \item We build a discrete event simulation
        for G/G/R-JSQ-PS systems;
    \item We propose six analytical approximations
        for the sojourn time cumulative distribution
        function (CDF) of M/M/R-JSQ-PS systems;
    \item We derive a run-time complexity analysis
        to obtain the sojourn time CDF using
        both the simulation and analytical
        approximations;
    \item We study which approximation is more
        accurate depending on the system load
        and number of CPUs;
    \item We study the accuracy of the best
        approximation for the latency percentiles
        required by URLLC services, i.e., from
        the 99\textsuperscript{th} percentile
        to the 99.999\textsuperscript{th}
        percentile.
\end{itemize}
In terms of Wasserstein distance,
the proposed analytical approximations deviate
less than a 2 out of 182 time units from the
sojourn time CDF in M/M/R-JSQ-PS systems.
For the 99.99\textsuperscript{th} percentile,
the best approximation
yields an error of less than 1.78~time units.


The paper is structure as follows.
In Section~\ref{sec:mmr-jsq-ps} we introduce
the considered system that we study in this paper.
Then, in~Section~\ref{sec:related} we go over the related
work about the sojourn time CDF in queueing
systems. Later, in~Section~\ref{sec:simulator} we
discuss the development of the G/G/R-JSQ-PS simulation,
and in~Section~\ref{sec:analytical}
we detail the
analytical approximations that we propose for the
sojourn time CDF of M/M/R-JSQ-PS systems. Afterwards,
in~Section~\ref{sec:complexity} and
Section~\ref{sec:comparison}
we study the run-time complexity and accuracy
of the proposed approximations, respectively.
Then, in Section~\ref{sec:behaviour-high-reliabilities}
we study the accuracy of our approximations
at the percentiles required by URLLC services.
Finally,
in~Section~\ref{sec:conclusions}
we conclude our work and 
point out future research directions.


\section{An M/M/R-JSQ-PS queueing system}
\label{sec:mmr-jsq-ps}

This work is concerned with the sojourn time distribution
$\mathbb{P}(T \leq t)$ of customers in an M/M/R-JSQ-PS system,
that is a system with $R$ parallel processor-sharing queues,
with overall Poisson arrival rate $\Lambda$, and intended
service times distributed exponentially with rate $\mu$.
Customers join the processor-sharing queue that has the
least amount of customers.

Processor-sharing (PS) is a queueing discipline where all
customers are served simultaneously, but the service load is
shared between the customers. That is, if a customer is
expecting to receive a service time $s$, then the rate at
which that service is given is $s/n$ when there are $n$
customers present. Therefore if there are $n$ customers
present throughout the customer's service, then it will last
$sn$ time units. A key feature is that $n$ can vary during
that customer's service.

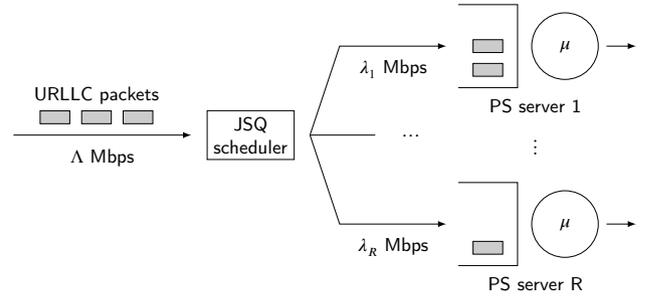
\begin{figure}[t]
    \centering
    \resizebox{\columnwidth}{!}{%
        \input{mmc.tex}
    }
    \caption{M/M/R-JSQ-PS system processing URLLC packets.}
    \label{fig:mmrjsqps}
\end{figure}

Figure~\ref{fig:mmrjsqps} illustrates this system.
Note that URLLC packets can be considered as
customers in the context of queueing theory,
hence, throughout the paper
we refer to customers as it a standard term in
queuing theory.

\section{Related work}
\label{sec:related}

In the networking community queuing theory
is a well-established tool to assess the
modelling of network
infrastructure~\citep{kleinrock,mor}.
The packet-based nature of the Internet,
so as the buffering and processor-sharing
nature of servers, make it a useful theoretical
tool to derive insights on the behaviour
of the network. Recent URLLC services and their
urgent need for communication guarantees can
benefit from the theoretical results of queuing
theory in order to adequately evaluate the network performance.

The fundamental results
of queuing theory~\citep{kleinrock}
give closed-form formulas for the sojourn
time (waiting plus service time)
of M/M/1 systems, i.e. systems
with 1 server that has exponential service
time to dispatch customers arriving
according to a Poisson distribution
and queue up before they are served.
Namely, it is possible to find both
the average and CDF for the sojourn time
of M/M/1 systems, with the latter having
also an exponential
distribution~\citep{kleinrock}.

However, the internet traffic is typically
dispatched in parallel by multiple servers
or CPUs within a server. Hence, it is
better resorting to M/M/R systems with
up to $R$ servers (or CPUs) that attend customers
in parallel. For such systems, the
queueing theory fundamentals also give
closed-form expressions for the
average sojourn time~\citep{kleinrock},
and indications on how to derive
its CDF~\citep{mor}.

But still, both M/M/1 and M/M/R systems
may not be suitable to model networking
components. Either because the assumption
of Poissonian arrivals is not suitable
or because considering
exponential service times is not
realistic. To that end, the literature
has devoted effort to derive the
sojourn time CDF expressions of systems
not satisfying such assumptions.
For example \citep{md1,mg1} provide expressions
for the sojourn time CDFs of M/D/1
and M/G/1 systems, respectively. However
both works provide the sojourn
time CDF expression in the form of the
Laplace-Stieltjes transform, i.e. a
non-closed expression of the sojourn
time CDF. Other works such as~\citep{masuyama2003sojourn}
shift the interest to systems that
follow Markovian arrival processes (MAP),
rather than Poissonian and provide
closed-formulas for the sojourn time CDF
in MAP/M/1 systems.

In general, making the assumption of
Poissonian arrivals is fair as long
as there is a considerable amount of
independent flows, as the
Palm–Khintchine theorem
states~\citep{palm}.
Hence, it is reasonable to model 
data centres as M/G/R systems,
as suggested by~\citep{mor}.
Namely, \citep{mor}~motivates the study of
M/M/R systems as server farms for traffic
processing, and the book leaves as exercise
how to derive the sojourn time CDF
of an M/M/R system
following the strategy used for M/M/1 systems.
Nevertheless, M/M/R systems do not mimic
the behaviour of Linux based systems
where each CPU shares the computing time 
using a processor-sharing discipline, rather
than the one-at-a-time processing of
M/M/R, where packets wait in the queue
until a server finishes processing a job.
As such, \citep{sqa}~propose to model
web server farms using
M/M/R-JSQ-PS systems with jobs joining
the CPU with the shortest queue (JSQ),
and each CPU serving all it's jobs
simultaneously via a processor-sharing
discipline (here joining the `shortest
queue' implies joining the CPU with the
smallest current load). The
research resorts to single queue analysis
(SQA) to provide insights on how the
traffic intensity changes depending on
the queue occupation at each CPU,
so as the average number of jobs at each
CPU.

\bigskip
The queuing theory
literature has widely studied the sojourn time
in different systems, and has managed to
find out not only their average sojourn time
but also the CDF.
However, the latter has only been possible
in some systems that do not capture
the multiple CPUs with PS fashion of
Linux based servers.
To the best of our knowledge, the
existing literature does not provide
expressions to compute the sojourn time
CDF in PS multi-processor systems that
are close to those servers that will
process URLLC traffic.
Therefore this paper contributes
to the related work by proposing six
approximations for the sojourn time CDF
of M/M/R-JSQ-PS systems. The proposed
approximations are useful to check whether
the URLLC traffic processing will meet
the 99\% or similar guarantees
of URLLC with almost negligible latencies
in the order of 1-10~milliseconds.

To check the accuracy of the proposed approximations
we resort to stochastic simulations of the
M/M/R-JSQ-PS system.
%
Discrete-event simulation is a
standard technique for the task \citep{robinson2014simulation},
with a number of commercial (e.g. Simul8
\citep{simul8} and AnyLogic \citep{anylogic})
and open-source (e.g. Simmer \citep{simmer},
SimPy \citep{simpy}, and Ciw \citep{palmer2019ciw})
software options.
However, to the authors' knowledge, prior to the
work of this paper the listed options do not offer
straightforward out-of-the-box ways to simulate
processor-sharing servers, requiring bespoke code
or modifications.
Therefore, another major contribution of this paper is
the extension of the Ciw software to be able to
simulate various kinds of processor-sharing queues.
This work is described in Section~\ref{sec:simulator}.

\section{Simulation of G/G/R-JSQ-PS}\label{sec:simulator}
In discrete event simulation a virtual representation of a
queueing system is created, and `run' by sampling a number
of basic random variables such as arrival dates of customers
and intended service times, which interact with one another
and the system to emulate the behaviour of the queueing
system under consideration. Given a long enough runtime and/or
a large enough number of trials, observed statistics will
converge to exact values due to the law of large numbers.
However due to their stochastic nature convergence may be
slow, and depending on the complexity of the system, can be
computationally expensive.
Here the Ciw library \citep{palmer2019ciw} is used, an
open-source Python library for conducting discrete event
simulation.
A key contribution of this work is the adaption of the library
to include processor-sharing capabilities, which were included
in release v2.2.0: these capabilities include standard processor-sharing,
limited processor-sharing as described in
\citep{zhang2009law}, and capacitated processor-sharing as
described in \citep{li2011radio}, and their combinations.

\begin{figure}
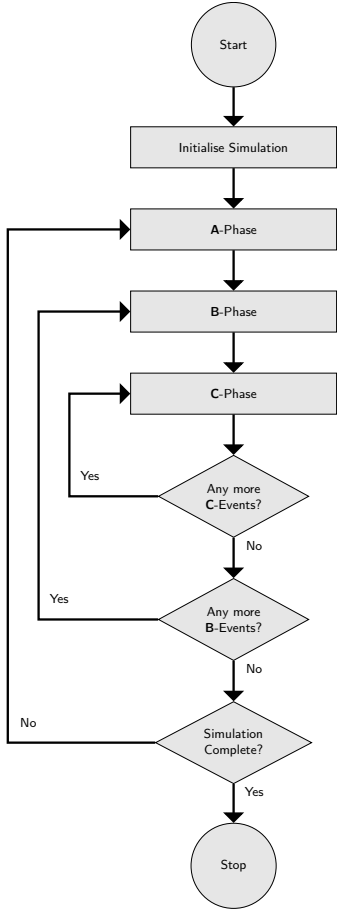

    \centering
    \includestandalone[width=0.25\textwidth]{img/eventschedulingapproach}
    \caption{Flow diagram of the event scheduling approach used by Ciw, taken from \citep{palmer18}.}
    \label{fig:eventscheduling}
\end{figure}

Ciw uses the event-scheduling approach to discrete event
simulation \citep{palmer2019ciw}.
Here time jumps from event to event in a discrete manner, while
events themselves can cause any number of other events to be
scheduled, either immediately or at some point in the future.
If they are scheduled for the future, then they are called
\textbf{B}-events, for example the event of a customer beginning
service will cause a future scheduled event of that customer
finishing service. If the events are scheduled immediately, then
they are called conditional or \textbf{C}-events, for example the
event a customer joining a queue may immediately cause another
event, that customer beginning service, if there was enough service
capacity. In addition to scheduling events, events can cause future
events to be re-scheduled for a later or earlier time.
A \textbf{B}-event, and its scheduling and re-scheduling of future
events, is called the \textbf{B}-phase; a \textbf{C}-event, and its
scheduling and re-scheduling of future events is called the
\textbf{C}-phase; and advancing the clock to the next
\textbf{B}-event is called the \textbf{A}-phase.
Figure~\ref{fig:eventscheduling} illustrates this event
scheduling process.

Processor sharing is implemented by manipulating the re-scheduling
of future events in the following way.
Upon arrival, a customer is given an arrival date $t_\star$, and
an intended service time $s$. They also observe the number of
customers, including themselves, who are present at the
processor-sharing server, $x_\star$.
At this point they have already received $d = 0$ of their
intended service time.
Given that nothing else changes, this customer will finish
service at date $t_{\text{end}}$ calculated
in \eqref{eqn:reschedule}.

\begin{equation}\label{eqn:reschedule}
t_{\text{end}} = t_\star + \frac{1}{x_\star} (s - d)
\end{equation}

Therefore this is the date that will be scheduled for that
customer to finish service.
Now, say and event happens at some $t$ such that
$t_\star < t < t_{\text{end}}$, and that event is either an
arrival to the server, or another customer finishing service
with the server. If the event is an arrival, set
$x = x_\star + 1$; and if the event is a customer finishing
service then set $x = x_\star - 1$. At this point our original
customer will have received
$d = d + \frac{1}{x_\star}(t - t_\star)$ of their intended
service. Now set $x_\star = x$, $t_\star = t$, and re-calculate
their service end date using \eqref{eqn:reschedule}, and then
re-schedule their finish service event.

This re-scheduling process is to be performed for every customer
in service at any \textbf{B}- or \textbf{C}- event that causes
$x_\star$ to change. This was implemented and released in
Ciw~v2.2.0, along with some processor-sharing variations: ($i$) limited
processor-sharing queues \citep{zhang2009law}, a generalisation of
a processor-sharing queue, in which only a given number of customers
may share the service load at any one time; and
($ii$) capacitated processor-sharing
queues \citep{li2011radio} with a switching parameter, where
the service discipline flips from FIFO to processor-sharing
if the number of customers exceeds this parameter.

The join-shortest-queue processor-sharing system considered in
this paper -- see~Fig.~\ref{fig:mmrjsqps} --
is implemented by combining this processor-sharing
capability with custom routing (JSQ) using inheritance of Ciw's modules.
An example is given in the documentation:
\url{https://ciw.readthedocs.io/en/latest/Guides/behaviour/ps_routing.html}.
Sojourn time CDFs can then be calculated easily as all customer
records are saved, namely, the sojourn time
of each customer is derived as $t_{\text{end}}-t_*$.

\section{M/M/R-JSQ-PS sojourn time CDF approximations}
\label{sec:analytical}

\begin{table}[t]
    \caption{Notation table}
    \label{tbl:notation}
\centering
\begin{tabular}{ c l  }
\toprule
\textbf{Symbol} & \textbf{Definition} \\
\midrule
$T$ & random variable for the customer sojourn time \\
$\Lambda$ & overall arrival rate. \\
$\mu$ & intended service rate \\
$R$ & number of parallel processor-sharing servers \\
$\rho$ & traffic intensity $\rho = \frac{\Lambda}{R\mu}$ \\
$\lambda_n$ & arrival rate seen by a server with $n$ customers \\
$W$ & complementary sojourn time CDF: $\mathbb{P}(T > t)$ \\
$w_n$ & $W$ with $n$ customers, $w_n(t) = \mathbb{P}(T > t \;|\; n)$ \\
$A_n$ & probability of joining a server with $n$ customers \\
$\pi_n$ & portion arrivals at a server with $n$ customers \\
$C(\mathbf{v}, b)$ & number of occurrences of $b$ in the vector $\mathbf{v}$ \\
$Z(\mathbf{v}, b)$ & set of indices in $\mathbf{v}$ where $b$ occurs \\
$Q$ & system transition matrix, with entries $q_{i,j}$ \\
$p_j$ & probability of being in state $j$  \\
$D$ & defective infinitesimal generator  \\
$L_1$ & maximum number of customers at a server\\ 
$L_2$ & maximum number of customers at the system\\ 
$q_{\text{max}}$ & maximum runtime of the simulation \\
$q_{\text{warmup}}$ & warmup time used in the simulation \\
$t_{\text{max}}$ & largest value of $T$ calculated \\
$\Omega(G, H)$ & Wasserstein distance between CDFs $G$ and $H$ \\
\bottomrule
\end{tabular}
\end{table}

In order to find the sojourn time distribution of a join-shortest-queue processor-sharing M/M/R-JSQ-PS queue, we follow an approach outlined in \citep{sqa}, called Single Queue Analysis (SQA). Here, rather than consider the whole M/M/R queue, we consider each server as it's own M/M/1-PS queue, with state-dependent arrival rates dependent on the join-shortest-queue mechanism.
Let $\Lambda$ denote the overall arrival rate to the M/M/R-JSQ-PS system, then for each PS server
in~Fig.~\ref{fig:mmrjsqps}
their effective state-dependent arrival rate is $\lambda_n$ when there are $n$ customers already being served by that server.
Table~\ref{tbl:notation} summarizes the notation used throughout this
paper.

Now, considering a single server as its own queue,  we adapt the methodology developed in \citep{masuyama2003sojourn} to the join-shortest-queue situation. In that paper Theorem 1 gives the sojourn time CDF of a single MAP/M/1-PS queue. A small adaptation, now considering an generic MAP process state-dependent Markovian arrivals $\lambda_n$, gives the sojourn time CDF as:

\begin{equation}\label{eqn:sojourn_time_cdf}
    \mathbb{P}(T \leq t) = 1 - \mathbb{P}(T > t) = 1 - W(t) = 1 - \sum_{n=0}^{\infty} A_n w_n(t)
\end{equation}
where $A_n$ is the probability of an arriving customer joining the queue when there are $n$ customers already present, and $w_n(t)$ is the conditional probability that the sojourn time is greater than $t$ given that there are $n$ customers already present at arrival.

We study two approximations each for finding the $\lambda_n$, $A_n$, and $w_n$ for each $n$. Then combining these in \eqref{eqn:sojourn_time_cdf} gives us six approximations of the sojourn time CDF for an M/M/R-JSQ-PS queue.

\subsection{First approximation of $\lambda_n$}\label{sec:lambdan_mc}
Note first that the arrival rate for each single queue being dependent on the number of customers already present in that queue is a valid assumption: the arrival rates to each individual queue when there are $n$ customers already present depends on the probability of $n$ being the smallest number of customers present in all $R$ of the queues. This however is not straightforward to calculate in isolation of the other $R$ queues, therefore we resort to approximations.
 
First we note that $\lambda_n = \pi_n \Lambda$, where $\pi_n$ is  the proportion of arrivals a server will receive if they have $n$ customers already present.

\begin{figure}
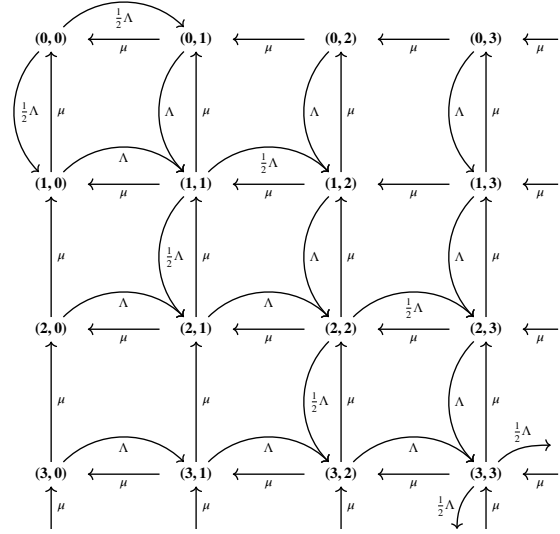

\begin{center}
\includestandalone[width=\columnwidth]{markov_chain}
\end{center}
\caption{Transition state diagram of the M/M/R-JSQ-PS system when $R=2$.}
\label{fig:markovchain}
\end{figure}

We find $\pi_n$ by constructing a truncated Markov chain of the M/M/R-JSQ-PS system. Define the state space of the non-truncated Markov chain by
\begin{equation}
    \label{eq:states-markov}
    S = \{(a_1, a_2, \dots, a_R) \; \forall \; a_1, a_2, \dots, a_R \in \mathbb{N}_0\}
\end{equation}
where $a_z$ denotes the number of customers with server $z$. Order the states and let $\mathbf{s}_i$ be the $i$th state. Define the transition rate $q_{i, j}$ from $\mathbf{s}_i$ to $\mathbf{s}_j$, for all $i$, $j$, by \eqref{eqn:transitions}:

\begin{equation}\label{eqn:transitions}
  q_{i, j} = \left\{
  \begin{matrix*}[l]
      \mu & \text{if } C(\delta, 0) = R-1 \land C(\delta, -1) = 1; \\
      \frac{\Lambda}{C(\mathbf{s}_i, \min(\mathbf{s}_i))} & \text{if } \delta = C(\delta, 0) = R-1 \land C(\delta, 1) = 1\\ & \land\ Z(\delta, 1) \subseteq Z(\mathbf{s}_i, \min(\mathbf{s}_i)); \\
      0 & \text{otherwise,}
  \end{matrix*} \right.
\end{equation}
where $\delta = \mathbf{s}_i - \mathbf{s}_j$; $C(\mathbf{v}, b) = |\{z \in \mathbf{v} : z = b\}|$ is a function that counts the number of occurrences of $b$ in a vector $\mathbf{v}$; and $Z(\mathbf{v}, b) = \{z : \mathbf{v}_z = b\}$ is the set of indices in $\mathbf{v}$ where $b$ occurs. Figure~\ref{fig:markovchain} is a representation of the Markov chain when $R=2$. When $R=1$ this reduces to an M/M/1 (or equivalently M/M/1-PS) system, and it becomes difficult to represent this system when $R>2$.

Steady-state probabilities can be found numerically by truncating the Markov chain, that is choosing an appropriate $L_1$ such that $a_z < L_1$ for all servers $z$, and solving $\mathbf{p} Q = \mathbf{0}$ with $\mathbf{p} \mathbf{e} = 1$, where $Q$ is the transition matrix with entries $q_{i, j}$ and $\mathbf{e}$ is the vector of ones.

Once all $p_i$ are found, the proportion of arrivals a server will receive if it has $n$ customers already present, $\pi_n$, can be found using \eqref{eqn:props}:

\begin{equation}\label{eqn:props}
\pi_n = 
    \left(\sum_{\substack{\mathbf{s}_{j, 0} = n \\ \min{\mathbf{s}_j} = n}} \frac{p_j}{C(\mathbf{s}_j, n)}\right)
    \left(\sum_{\mathbf{s}_{j, 0} = n} p_j\right)^{-1}
\end{equation}
where $\mathbf{s}_{j, 0}$ represents the number of customers at the first server when in state $j$.

\subsection{Second approximation of $\lambda_n$}\label{sec:lambdan_approx}

The authors of \citep{sqa} provide numerical approximations for $\lambda_{0}, \lambda_{1}, \lambda_{2}$ in~\citep[Section 5]{sqa}, given in \eqref{eqn:approxlambda0}, \eqref{eqn:approxlambda1} and \eqref{eqn:approxlambda2}, and all other $\lambda_n$ for $n \geq 3$ by \eqref{eqn:approxlambdan}.

\begin{align}
\lambda_0 &= \mu \left(k_a - k_b k_c^R - k_d k_e^R\right) \label{eqn:approxlambda0} \\
\lambda_1 &= \frac{\mu \left(\rho^{R} - 1 + \frac{\mu \left(\rho - \rho^{R + 1}\right)}{\lambda_{0} \left(1 - \rho\right)}\right)}{\frac{\lambda_{2}}{\mu} - \rho^{R} + 1} \label{eqn:approxlambda1} \\
\lambda_2 &= \mu k_f k_g^R \label{eqn:approxlambda2} \\
\lambda_n &= \mu\left(\frac{\Lambda}{n\mu}\right)^n \label{eqn:approxlambdan}
\end{align}

with $k_a$, $k_b$, $k_c$, $k_d$, $k_e$, $k_f$ and $k_g$ defined by:

\begin{align}
k_a &= \frac{\rho}{(1-\rho)}\\
k_b &= \frac{-0.0263\rho^2+0.0054\rho+0.1155}{\rho^2-1.939\rho+0.9534}\\
k_c &= -6.2973\rho^4+14.3382\rho^3-12.3532\rho^2\nonumber\\
    & +6.2557\rho-1.005\\
k_d &= \frac{-226.1839\rho^2+342.3814\rho+10.2851}{\rho^3-146.2751\rho^2-481.1256\rho+599.9166)}\\
k_e &= 0.4462\rho^3-1.8317\rho^2+2.4376\rho-0.0512\\
k_f &= -0.29 \rho^3 + 0.8822 \rho^2 - 0.5349 \rho + 1.0112\\
k_g &= -0.1864 \rho^2 + 1.195 \rho - 0.016
\end{align}

\subsection{First approximation of $A_n$}\label{sec:An_mc}

Using the same Markov chain defined in Section~\ref{sec:lambdan_mc}, $A_n$ can be found by manipulating the steady-state probabilities $p_n$, given in \eqref{eqn:An}:

\begin{equation}\label{eqn:An}
    A_n = \sum_{\min{\mathbf{s}_j} = n} p_j.
\end{equation}

\subsection{Second approximation of $A_n$}\label{sec:An_approx}

From the SQA we can consider each PS server to be its own M/M/1-PS queue with state-dependent arrival rates. This gives a birth-death process, where $A_n$ is the probability of that system being in state $n$. Thus we have:

\begin{align}
    A_n &= \prod_{i=0}^{n-1} \frac{\lambda_i}{\mu} A_0\label{eqn:An2}\\
    A_0 &= \left( 1 + \sum_{i=1}^{\infty} \prod_{j=0}^{i-1} \frac{\lambda_j}{\mu} \right)^{-1}.\label{eqn:A0}
\end{align}

\subsection{First approximation of $w_n(t)$}\label{sec:wnt_matrix}

Again, we resort to SQA and focus on one
server of our M/M/R-JSQ-PS system in
Fig.~\ref{fig:mmrjsqps}. As aforementioned, such
server behaves as an M/M/1-PS queue with
state-dependant arrivals at rate $\lambda_n$,
and has a complementary sojourn time CDF $w_n(t)$
when it is attending $n$ customers.

We follow the strategy
from~\citep[Section 3]{masuyama2003sojourn}, where
authors derive $w_n(t)$ for an MAP/M/1-PS queue.
Specifically, we derive
$\mathbf{w}(t) = (w_0(t), w_1(t), w_2(t),\ldots)$
as the solution of the differential equation
$\tfrac{d}{dt}\mathbf{w}(t)=D \mathbf{w}(t)$,
which is:
\begin{equation}
    \mathbf{w}(t) = e^{D t} \mathbf{e}
    \label{eqn:exp-generator}
\end{equation}
with $D$ the defective infinitesimal
generator for our state-dependant M/M/1-PS queue
in the SQA:
 \begin{equation}\label{eqn:defective_IG}
 \resizebox{0.85\hsize}{!}{%
     $D = \begin{pmatrix}
         -(\lambda_0+\mu) & \lambda_0 & 0 & 0 & \ldots \\
         \frac{1}{2}\mu & -(\lambda_1+\mu) & \lambda_1 & 0 & \ldots \\
         0 & \frac{2}{3}\mu & -(\lambda_2+\mu) & \lambda_2 & \ldots \\
         0 & 0 & \frac{3}{4}\mu & -(\lambda_3+\mu) & \ldots \\
         \vdots & \vdots & \vdots & \vdots & \ddots
     \end{pmatrix}$%
 }
 \end{equation}


By constructing a truncated $D$ explicitly, numerical methods,
such as Padé's method~\citep{pade},
are used to find the matrix exponential.
As a result, $w_n(t)$ is obtained as the
$n$\textsuperscript{th} entry of the
$\mathbf{w}(t)$ vector 
in~\eqref{eqn:exp-generator}.

\subsection{Second approximation of $w_n(t)$}\label{sec:wnt_unroll}

As constructing $D$ explicitly and numerically computing a matrix exponential can be computationally inefficient, in Lemma~\ref{lemma:soj-conditioned} we give a recurrent relation for finding $w_n(t)$.

\begin{lemma}
    \label{lemma:soj-conditioned}
    If a server within an M/M/R-JSQ-PS system has $n$
    customers, its sojourn time CDF is
    \begin{equation}
        \mathbb{P}(T > t\;|\;n) = w_n(t) =
        \sum_{i=0}^{\infty}\frac{(\lambda_{0} + \mu)^i t^i}{i!} e^{-(\lambda_{0} + \mu)t} h_{n, i}
        \label{eq:soj-conditioned}
    \end{equation}
    with $h_{n, 0} = 1$ for all $n$, $h_{-1, i} = 0$ for all $i$, and $h_{n,i}$
    satisfying
    \begin{multline}
        h_{n, i+1} = \frac{n}{n+1}\frac{\mu}{\lambda_{0} + \mu} h_{n-1, i} + h_{n,i}\left(1 - \frac{\lambda_n+\mu}{\lambda_{0}+\mu}\right)\\
        + \frac{\lambda_n}{\lambda_{0} + \mu} h_{n+1, i}\label{eq:hs}
    \end{multline}
\end{lemma}
\begin{proof}
    We mimic the proof presented in~\citep[Corollary 2]{masuyama2003sojourn}.
    and apply the uniformisation technique~\citep{uniformization}
    for the matrix exponential
    in \eqref{eqn:exp-generator}. As a result
    we obtain:
    \begin{equation}
        \mathbf{w}(t)=\sum_{i=0}^{\infty}\frac{(\lambda_0 + \mu)^i t^i}{i!} e^{-(\lambda_0 + \mu)t} \left[ I + \frac{1}{\lambda_{0}+\mu}D \right]^i \mathbf{e}
    \end{equation}
    with $I$ the identity matrix. To ease the computation
    of the matrix to the power of $i$, (i.e., $[\cdot]^i$)
    the following vector is defined
    $\mathbf{h}_{n,i}=\left[I+\tfrac{1}{\lambda_{0}+\mu}D\right]^i\mathbf{e}$.
    And it leads to the recursion $\mathbf{h}_{n,i+1}=\left[I+\tfrac{1}{\lambda_{0}+\mu}D\right]\mathbf{h}_{n,i}$,
    with $\mathbf{h}_{n,0}=\mathbf{e}, \forall n$.
    As a result, $\mathbf{w}(t)$ is defined as
    \begin{equation}
        \mathbf{w}(t) = \sum_{i=0}^\infty \frac{(\lambda_{0} + \mu )^i t^i}{i!} e^{-(\lambda_{0}+\mu)t} \mathbf{h}_{n,i}
    \end{equation}
    and the n\textsuperscript{th} element of $\mathbf{w}(t)$ is given
    by~\eqref{eq:soj-conditioned}.
\end{proof}

This gives $w_n(t)$ in a form which, for a sufficiently large value, $L_2$, in place of infinity, can be found recursively. This naive adaptation of \citep{masuyama2003sojourn} replaces their static MAP with the state-dependent arrival rate $\lambda_n$.

\subsection{Summary \& Considerations}
\label{subsubsec:considerations}

In this work we implement and test six different methods of approximating the complementary sojourn time CDF of an M/M/R-JSQ-PS system, $W(t)$. Table~\ref{tbl:methods} summarises the methodology.

\begin{table}
    \centering
    \caption{Summary of the six methods of calculating $W(t)$.}
    \begin{tabular}{cccc}
      \toprule
      Method & $\lambda_n$ & $A_n$ & $w_n(t)$ \\
      \midrule
      \textbf{A} & \ref{sec:lambdan_mc} & \ref{sec:An_mc} & \ref{sec:wnt_matrix} \\
      \textbf{B} & \ref{sec:lambdan_mc} & \ref{sec:An_approx} & \ref{sec:wnt_matrix} \\
      \textbf{C} & \ref{sec:lambdan_approx} & \ref{sec:An_approx} & \ref{sec:wnt_matrix} \\
      \textbf{D} & \ref{sec:lambdan_mc} & \ref{sec:An_mc} & \ref{sec:wnt_unroll} \\
      \textbf{E} & \ref{sec:lambdan_mc} & \ref{sec:An_approx} & \ref{sec:wnt_unroll} \\
      \textbf{F} & \ref{sec:lambdan_approx} & \ref{sec:An_approx} & \ref{sec:wnt_unroll} \\
      \bottomrule
    \end{tabular}
    \label{tbl:methods}
\end{table}

Choices of model hyper-parameters, those that concern only the methodology and not the system that is itself being modelled, can effect both the accuracy and run-time (or computational complexity) of the model, and choices are usually a compromise between the two. For the Ciw simulation there are three hyper-parameters to consider: the maximum simulation time, the warm up time, and the number of trials. The larger the number of trials, the more we can smooth out the stochastic nature of the DES by take averages of the key performance indicators of each trial, however the more trials take longer to run. The warm-up time is a proportion of the maximum simulation time where results are not collected. This filtering of results ensures that key performance indicators are not collected before the simulation reaches steady-state, and therefore and not dependent on the starting conditions of the simulation. The larger the warm-up time, the higher the chance that the collected results are in steady state (this is highly dependent on other model parameters), although this means less results to collect and so more uncertainty. A larger maximum simulation time does both, ensures that there are enough results to decrease uncertainty, and increases the chance that steady-state is reached, however this also increases run-times.

Each of the six sub-methods described in Section~\ref{sec:analytical} have hyper-parameters than need to be chosen. Those that explicitly build an infinite Markov chain, that is methods~\ref{sec:lambdan_mc} and \ref{sec:An_mc}, need to truncate the Markov chain using a limit $L_1$, so that numerical methods can be used on a finite Markov chain. The limit $L_1$ corresponds to the maximum number of customers each PS server will receive. Thus these Markov chains will have $L_1^R$ states, and so its construction requires defining $L_1^{2R}$ transitions. The larger the $L_1$ the more accurate the model, as there would be a smaller probability of a server receiving more than $L_1$ customers, however larger limits have longer run-times and larger memory consumption.

Other sub-methods, methods~\ref{sec:An_approx} and~\ref{sec:wnt_unroll} contain infinite sums. For these, a sufficiently large cut-off, $L_2$ is required to truncate these sums for numerical computation. This $L_2$ corresponds to the overall maximum number of customers that can be present, and so can be chosen to much larger than $L_1$. Similarly, method~\ref{sec:wnt_matrix} requires the construction of a matrix, where each state corresponds the the overall number of customers, and so $L_2$ is also be used to truncate this matrix.

\subsection{Markov chain truncation}
When we approximate $\lambda_n$ and $A_n$
using Section~\ref{sec:lambdan_mc} and
Section~\ref{sec:An_mc}, respectively, we
truncate the transition matrix $Q$ of
the Markov chain in \eqref{eqn:transitions}.
Namely, we limit the ``last'' considered
state $S_i=(L_1-1)\mathbf{e}$ has $L_1-1$ users
in all the $R$ servers. The truncation
$L_1$ should be carefully selected such that
\begin{equation}
    \sum_{\mathbf{s}_j:\ \max \mathbf{s}_j\geq L_1}p_j <\varepsilon
    \label{eqn:truncation-tolerance}
\end{equation}
that is, the probability of entering a state
with a server with $L_1$ or more customers should
remain below a tolerance 
$\varepsilon\in\mathbb{R}^+$

\begin{figure}[t]
    \centering
    \includegraphics[width=\columnwidth]{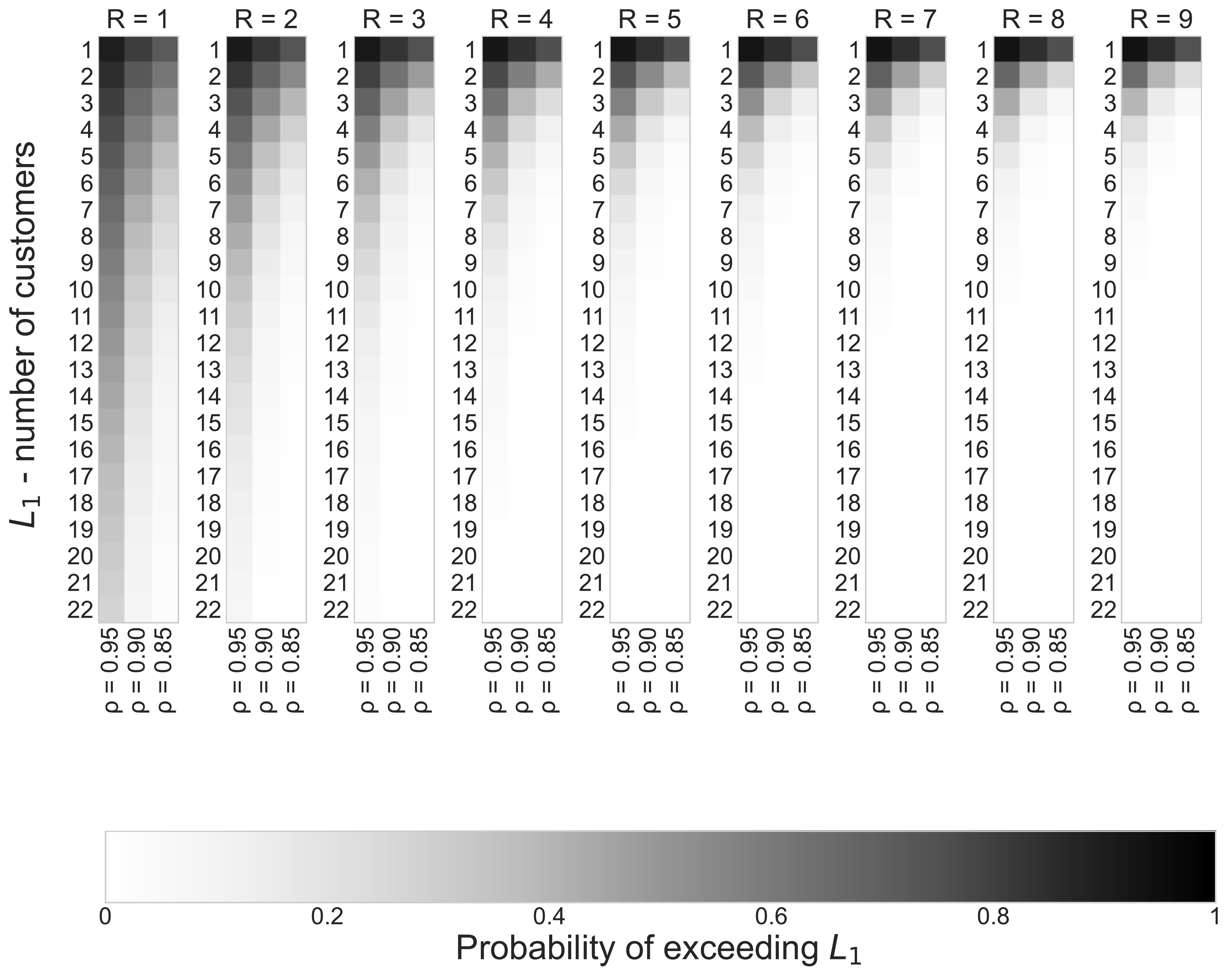}
    \caption{Probability of having
    $L_1$ or more customers at some server~\eqref{eqn:truncation-tolerance}
    with different loads $\rho=0.85,0.90,0.95$ and available servers $R=1,\ldots,9$.}
    \label{fig:mc-limit}
\end{figure}

Figure~\ref{fig:mc-limit} illustrates how
probability of having $L_1$ or more users at
a server decreases as we increase the
truncation limit $L_1$ and how this is
effected by both $\rho$ and $R$. This
data was obtained using the simulation
described in Section~\ref{sec:simulator}.

\section{Complexity analysis}
\label{sec:complexity}
As stated in the paper title, the main motivation
for modelling the M/M/R-JSQ-PS system is to tell
whether an URLLC service attended by a
multi-processor system meets latency and reliability
constraints. Hence, it is
of paramount importance to consider the run-time
complexity of each approximation $\lambda_n, A_n, w_n(t)$,
as a network operator may require fast operational
decisions to satisfy the URLLC. If the approximation
run-time is not fast enough, the operator would
not be able to update the operational decisions on
time upon demand changes --
e.g., increase the dedicated servers to attend
the increasing demand for an URLLC service.
Therefore,
in the following we analyse
the run-time complexity of each approximation for
$\lambda_n, A_n$ and $w_n(t)$.



\subsection{First approximation of $\lambda_n$}
Using~\eqref{eqn:props} this approximation finds
the portion of arrivals that a server foresees using
the steady-state probabilities $p_i$ of the
\mbox{M/M/R-JSQ-R} Markov chain with $L_1^R$ states
and transition matrix $Q$ of size $L_1^{2R}$.
For each entry $q_{i,j}$ of the transition matrix
we make $\min(\mathbf{s}_i),Z(\delta,b),C(\delta,b)$
operations, all of them of complexity $\mathcal{O}(R)$.
Hence, computing all entries of the transition matrix
$Q$ takes $\mathcal{O}(R L_1^{2R})$ operations.

To find the steady-state vector $\mathbf{p}$ we solve
$(\tilde{Q}|\mathbf{e})^T \mathbf{p}= (\mathbf{p}|\mathbf{e})^T$,
where $\tilde{Q}$ is the transition matrix $Q$ less one row.
This is a linear system with a matrix
of size $L_1^R \times L_1 ^ R$.
Finding such solution with
the LAPACK~\citep{lapack} \verb|gesv| method leads
to a cubic run-time complexity on the
matrix size. Therefore, obtaining the
steady-state probability has complexity
$\mathcal{O}\left(L_1^{3R}\right)$.
Note that it is the computation of $\mathbf{p}$
that dominates the complexity of approximating
$\lambda_n$, as creating the transition
matrix $Q$ has $\mathcal{O}\left(RL_1^R\right)$ complexity
and computing $\pi_n$
has $\mathcal{O}\left(L_1^R\right)$ complexity
-- see~\eqref{eqn:props}. Hence,
the first approximation of $\lambda_n$
has run-time complexity $\mathcal{O}\left(L_1^{3R}\right)$.

\subsection{Second approximation of $\lambda_n$}
In~\eqref{eqn:approxlambdan} we see that
there is a power relationship between $n$
and $\lambda_n$, namely,
$\lambda_n=\mu(\tfrac{\Lambda}{n\mu})^n$.
As computing a power has complexity
$\mathcal{O}(\log n)$, the second approximation of
$\lambda_n$ has complexity $\mathcal{O}(\log n)$.

    
\subsection{First approximation of $A_n$}
Once we compute the Markov chain
steady-state probabilities $p_n$,
this method only performs a summation
over such probabilities~\eqref{eqn:An}. Thus, the
complexity of computing $A_n$ is $\mathcal{O}\left(L_1^R\right)$,
for we iterate over all the $L_1^R$ states and
check whether each of them satisfies $\min \mathbf{s}_j=n$.

\subsection{Second approximation of $A_n$}
Given the values of $\lambda_n$, first
we compute the probability of joining
the queue when there are 0 users $A_0$
in~\eqref{eqn:A0}. As mentioned
in~Section~\ref{subsubsec:considerations}, we
truncate the infinite summations up to $L_2$.
Hence, it takes
$\sum_i^{L_2} i$ operations
to compute $A_0$, and so is $\mathcal{O}\left(L_2^2\right)$.
Once $A_0$ is computed, we perform
$\mathcal{O}(L_2)$ operations to compute
$A_n$ in~\eqref{eqn:An2}. Therefore as a whole,
the second approximation of $A_n$ has
$\mathcal{O}(L_2^2)$ complexity.

\subsection{First approximation of $w_n(t)$}
This approximation computes the exponential
of the defective infinitesimal generator
matrix $D$ -- see~\eqref{eqn:exp-generator}.
As mentioned in~Section~\ref{subsubsec:considerations},
we also truncate the $D$ matrix up to $L_2$
elements in its diagonal such that $D$ is
an $L_2 \times L_2$ matrix.
As $D$ is diagonal with $\leq3$ terms at each
row, its creation has complexity $\mathcal{O}(L_2)$.
With Padé's method~\citep{pade} we compute
$D$ exponential with $\mathcal{O}(L_2\log L_2)$
complexity.

\subsection{Second approximation of
$w_n(t)$}
Using the recurrent formula of
Lemma~\ref{lemma:soj-conditioned} we
can check the complexity of this second
approximation of $w_n(t)$.
As mentioned in~Section~\ref{subsubsec:considerations},
we truncate the infinite summation
in~\eqref{eq:soj-conditioned} to $L_2$
iterations. At each summation iteration $i$,
we perform $\mathcal{O}(\log i)$
operations (the power operators),
hence, computing the second approximation
of $w_n(t)$ has
complexity $\mathcal{O}(L_2 \log L_2)$.
Note that we compute $h_{n,i}$ incrementally
thanks to the recursive approach, hence,
such computation does not dominate the
approximation complexity as
$h_{n,i+1}$ reuses already computed
values of $h_{*,i}$. 
Similarly, we also keep inside a hash
table the factorial computations $i!$
at~\eqref{eq:soj-conditioned} denominator
to ease the computational burden.

\bigskip
Depending on which Method we use
-- see Table~\ref{tbl:methods} --
we will get different run-time complexities.
Namely, methods A,B,D, and E have
an $\mathcal{O}\left(L_1^{3R}\right)$ complexity because
they rely on the truncated Markov chain
to derive $\lambda_n$, which is the most
demanding approximation. While methods
C and F have an overall complexity of
$\mathcal{O}\left(L_2^2\right)$ because the
\ref{sec:lambdan_approx} and
\ref{sec:wnt_unroll} approximations dominate
the computation of $W(t)$. Table~\ref{tbl:complexity}
summarises the computational complexity of
each method. 

{\renewcommand{\arraystretch}{1.3}
\begin{table}
    \centering
    \caption{Complexity of each method.}
    \begin{tabular}{cllll}
      \toprule
      Method & $\lambda_n$ & $A_n$ & $w_n(t)$ & Overall \\
      \midrule
      \textbf{A} & $\mathcal{O}\left(L_1^{3R}\right)$ & $\mathcal{O}\left(L_1^{R}\right)$ & $\mathcal{O}\left(L_2\log L_2\right)$ & $\mathcal{O}\left(L_1^{3R}\right)$\\
      \textbf{B} & $\mathcal{O}\left(L_1^{3R}\right)$ & $\mathcal{O}\left(L_2^2\right)$ & $\mathcal{O}\left(L_2\log L_2\right)$ & $\mathcal{O}\left(L_1^{3R}\right)$\\
      \textbf{C} & $\mathcal{O}\left(\log n\right)$ & $\mathcal{O}\left(L_2^2\right)$ & $\mathcal{O}\left(L_2\log L_2\right)$ & $\mathcal{O}\left(L_2^2\right)$\\
      \textbf{D} & $\mathcal{O}\left(L_1^{3R}\right)$ & $\mathcal{O}\left(L_1^{R}\right)$ & $\mathcal{O}\left(L_2 \log L_2\right)$ & $\mathcal{O}\left(L_1^{3R}\right)$\\
      \textbf{E} & $\mathcal{O}\left(L_1^{3R}\right)$ & $\mathcal{O}\left(L_2^2\right)$ & $\mathcal{O}\left(L_2 \log L_2\right)$ & $\mathcal{O}\left(L_1^{3R}\right)$\\
      \textbf{F} & $\mathcal{O}\left(\log n\right)$ & $\mathcal{O}\left(L_2^2\right)$ & $\mathcal{O}\left(L_2 \log L_2\right)$ & $\mathcal{O}\left(L_2^2\right)$\\
      \bottomrule
    \end{tabular}
    \label{tbl:complexity}
\end{table}
}


\subsection{Simulation}

Events, and more importantly the number of events in a run
of the simulation are random. Therefore we cannot have a
true complexity analysis, but we can say something about the
order of expected number of operations. In this section we
consider the average time complexity of the M/M/R-JSQ-PS
system.

We will consider number of operations per unit of simulation
time when in steady state. Assuming there are $M$ customers
in the system at steady-state, there are two types of
\textbf{B}-events that can take place in a given time unit,
arrivals, and customers ending service.

\begin{itemize}
    \item \textit{Arrivals}: there's an average of $\Lambda$ arrivals
    per time unit. At each arrival we need to check $R$ servers
    to see which is least busy. Then once a server is chosen, we
    need to go through each customer at that server and re-schedule
    their end service dates - \eqref{eqn:reschedule}. As
    join-shortest-queue systems should evenly share customers between
    servers, we expect there to be $\frac{M}{R}$ customers at that
    server. So per time unit, the expected number of operations for
    arrival events is $\mathcal{O}\left(\Lambda\left(R + \frac{M}{R}\right)\right)$.
    
    \item \textit{End services}: at steady state, due to work conservation
    and Burke's theorem \citep{burke1956output}, there's an
    average of $\Lambda$ services ending per time unit. At each
    end service we need to go through each customer at that server
    and re-schedule their end service dates. So per time unit,
    the expected number of operations for end service events
    is $\mathcal{O}\left(\Lambda \frac{M}{R}\right)$.
\end{itemize}

It is difficult to find a closed expression for $M$, hence the
need for simulation and approximations. However a naive
estimate for the average number of customers $M$ is
the traffic intensity, $M \approx \rho = \frac{\Lambda}{\mu R}$.
Let $q_{\text{max}}$ be the maximum simulation time. Altogether,
in steady state the expected number of operations
is 
$\mathcal{O}\left(q_{\text{max}} \left( \Lambda \left( R + \tfrac{M}{R}\right) + \Lambda \tfrac{M}{R} \right) \right)$
for a simulation run,
which is equivalent to $ \mathcal{O}\left(q_{\text{max}} \left( \Lambda R^2 + \tfrac{2 \Lambda^2}{\mu R^2}\right) \right)$.

Although $q_{\text{max}}$ is a user chosen hyper-parameter,
and increases the expected number of operations linearly,
it is useful to consider if it's choice should be influenced
by other system parameters.
Consider that, when in steady state, increasing the simulation
time increases the number of sojourn time samples we have
to estimate the CDF. Say we need $X$ samples to estimate a
good CDF, then $q_{\text{max}}$ should be chosen such that
$q_{\text{max}} = \frac{X}{\Lambda}$. As $X$ is independent
of any other parameter, it can be considered a constant.
However, this is assuming a steady state. We should actually
choose $q_{\text{max}} = \frac{X}{\Lambda} + q_{\text{warmup}}$,
where $q_{\text{warmup}}$ is the warmup time, the time it takes
to reach steady state. It is likely that $q_{\text{warmup}}$
would be effected by the system parameters.

It is interesting to note that the six approximations' time
complexities, and the expected time complexity for the simulation,
are affected by different parameters. The approximations are
affected by the hyper-parameters $L_1$ and $L_2$, along with $R$,
however the simulation is effected by the system parameters
themselves. This shows that for some specific cases and parameter
sets, it might be worthwhile resorting to simulation after all.

\section{Approximations' accuracy}
\label{sec:comparison}

\begin{figure}
\begin{center}
\includestandalone[width=0.75\columnwidth]{wasserstein}
\end{center}
\caption{Graphical interpretation of the Wasserstein distance between the actual and approximated CDFs.}
\label{fig:wasserstein}
\end{figure}

\begin{figure*}
    \centering

    \subfloat[Method A]{\includegraphics[width=0.33\textwidth]{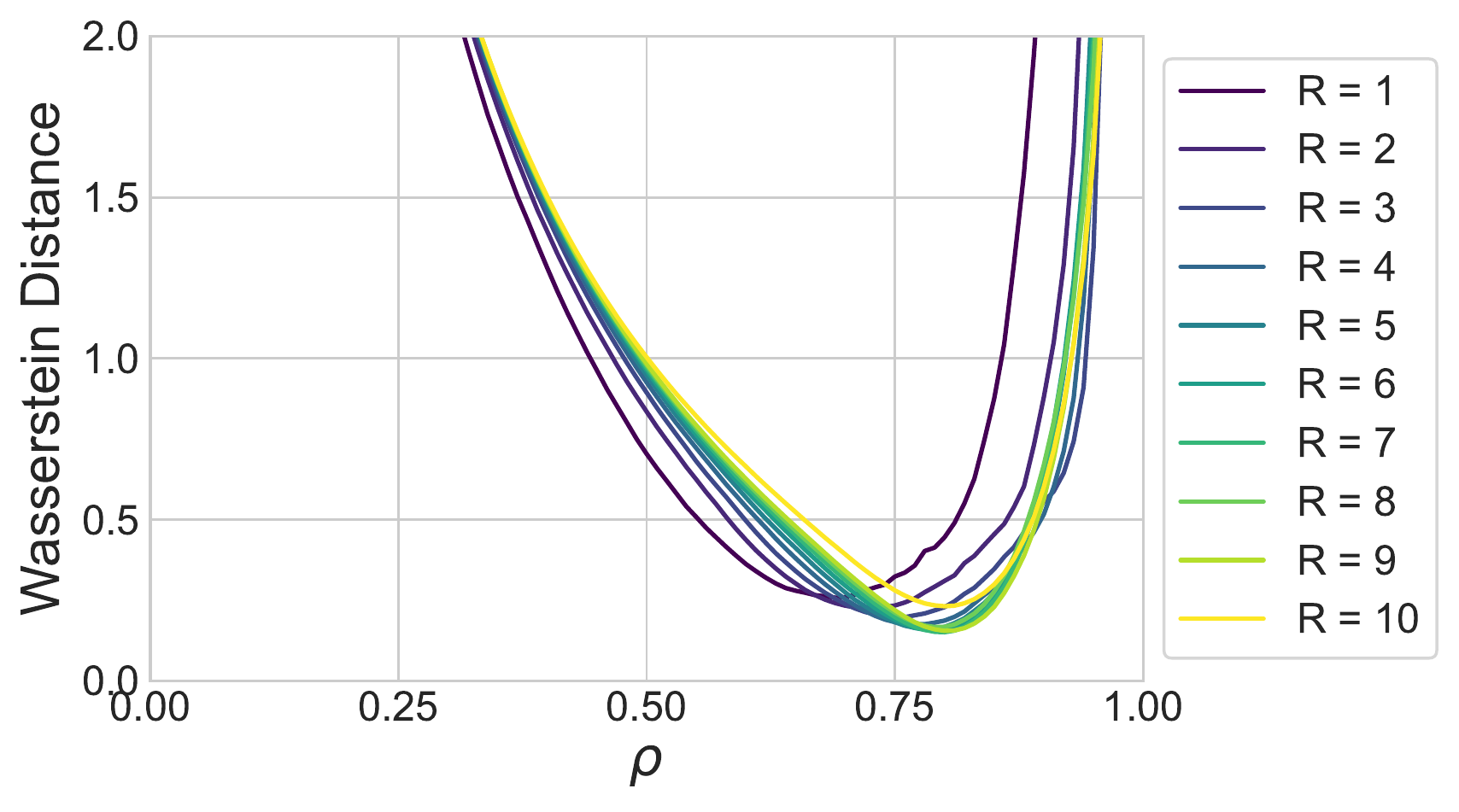}%
    \label{fig:accuracyA}}
    ~
    \subfloat[Method B]{\includegraphics[width=0.33\textwidth]{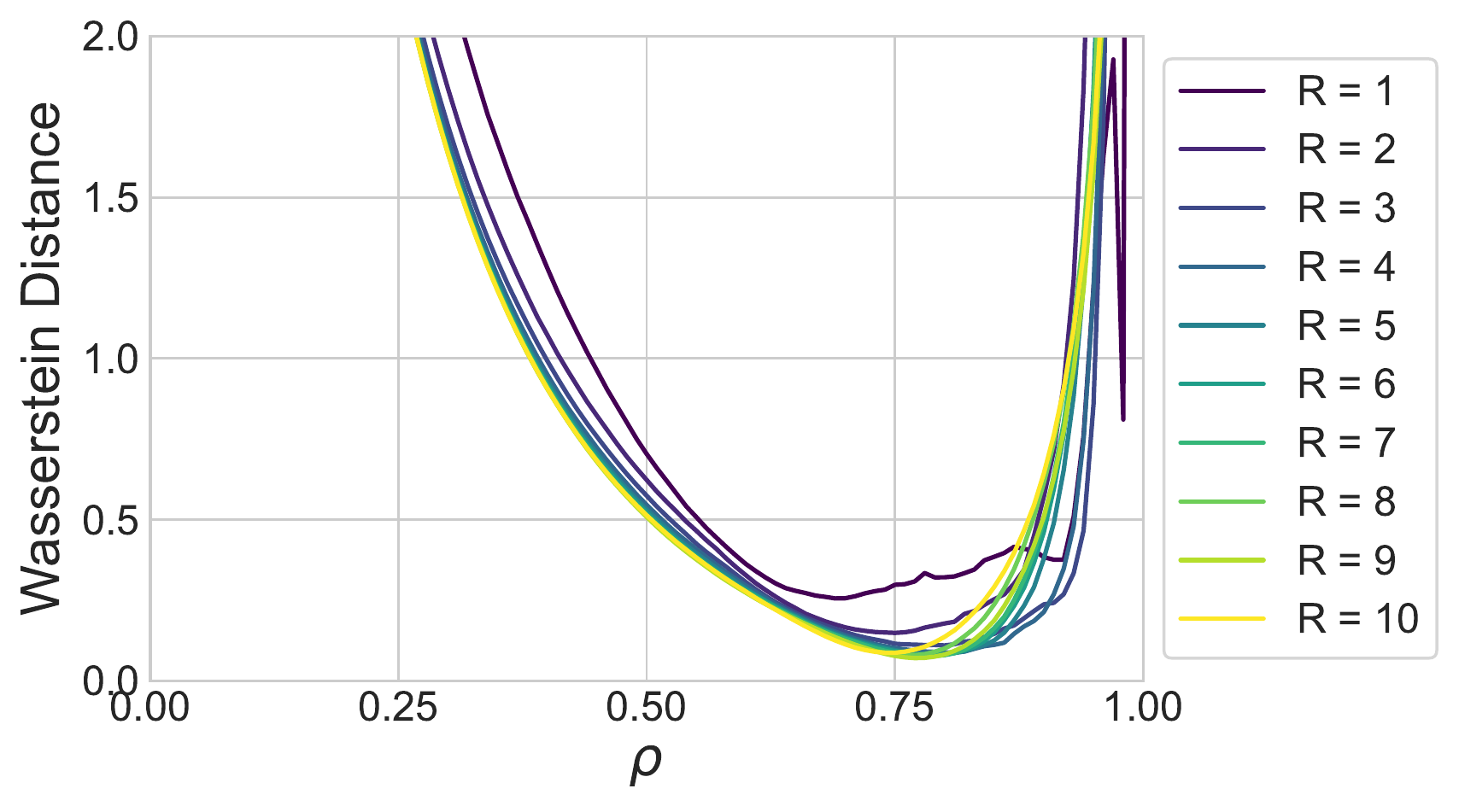}%
    \label{fig:accuracyB}}
    ~
    \subfloat[Method C]{\includegraphics[width=0.33\textwidth]{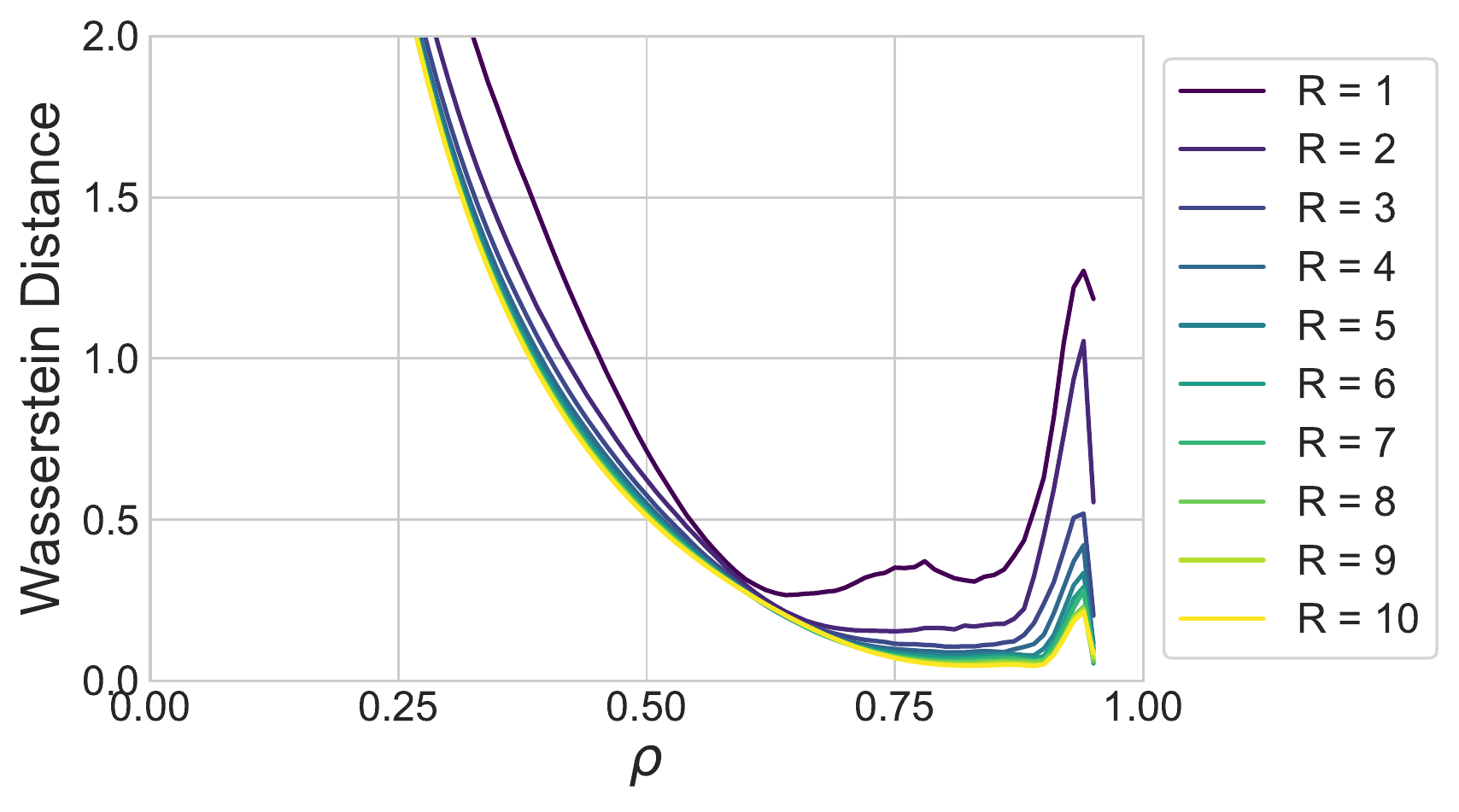}%
    \label{fig:accuracyC}}\\

    \subfloat[Method D]{\includegraphics[width=0.33\textwidth]{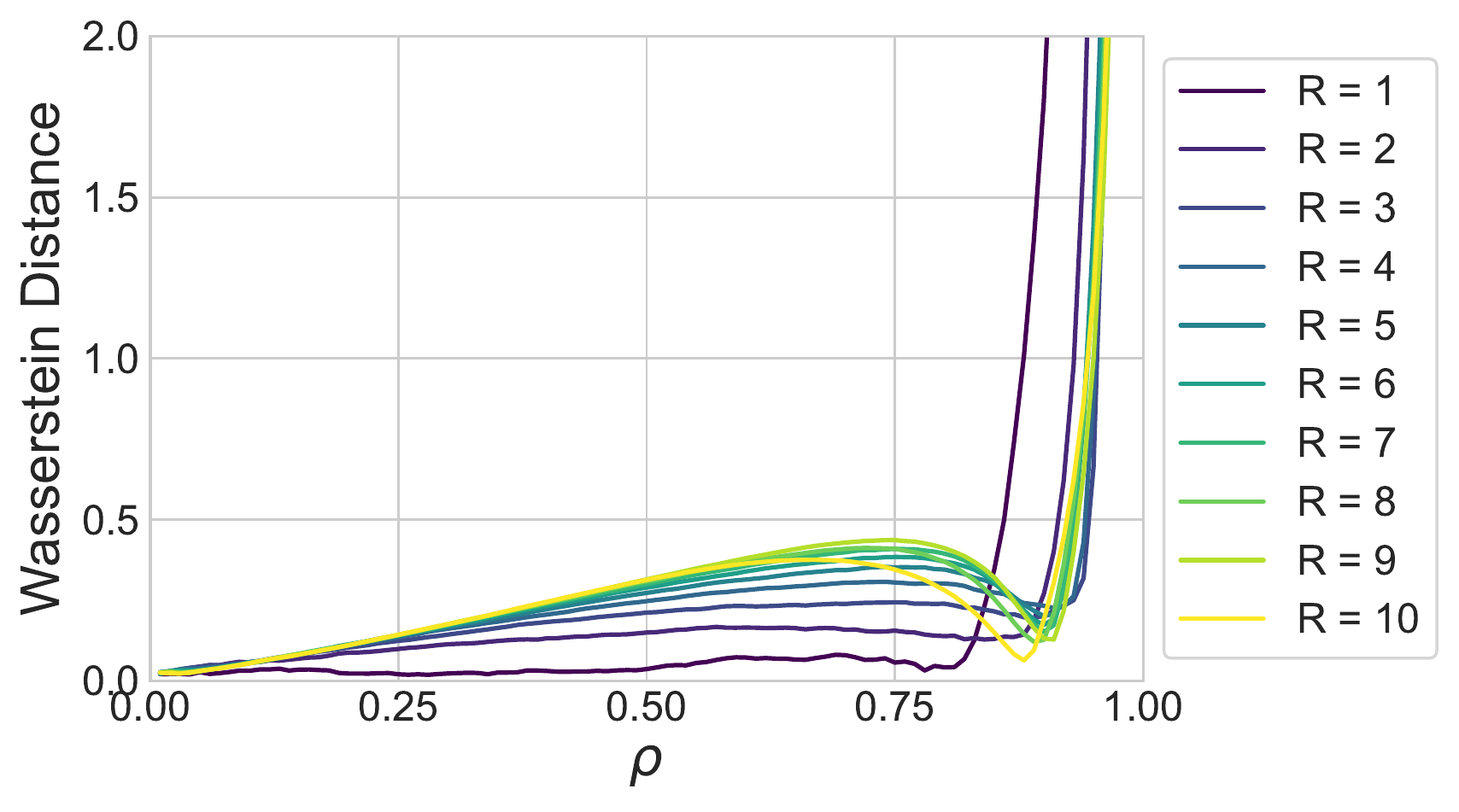}%
    \label{fig:accuracyD}}
    ~
    \subfloat[Method E]{\includegraphics[width=0.33\textwidth]{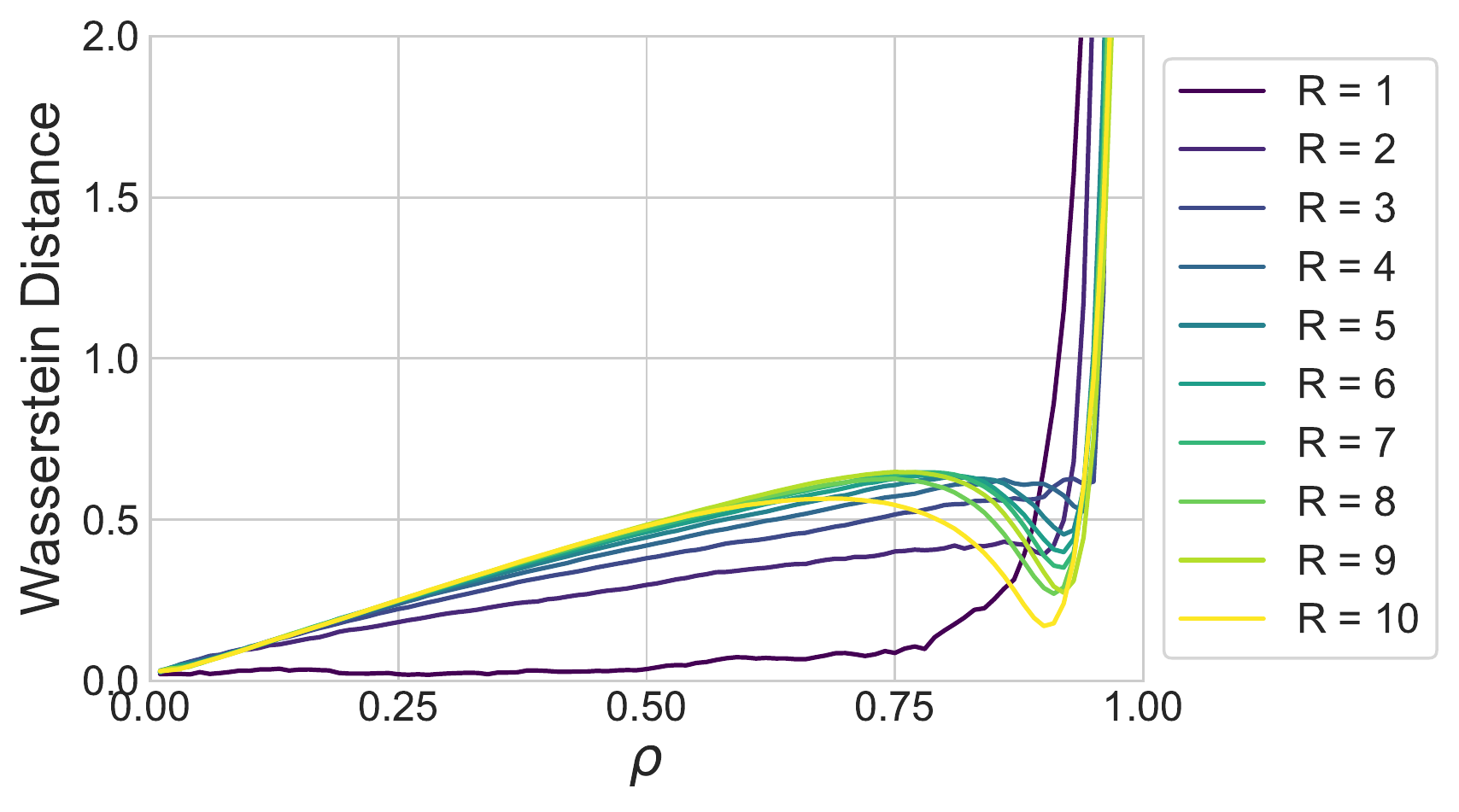}%
    \label{fig:accuracyE}}
    ~ 
    \subfloat[Method F]{\includegraphics[width=0.33\textwidth]{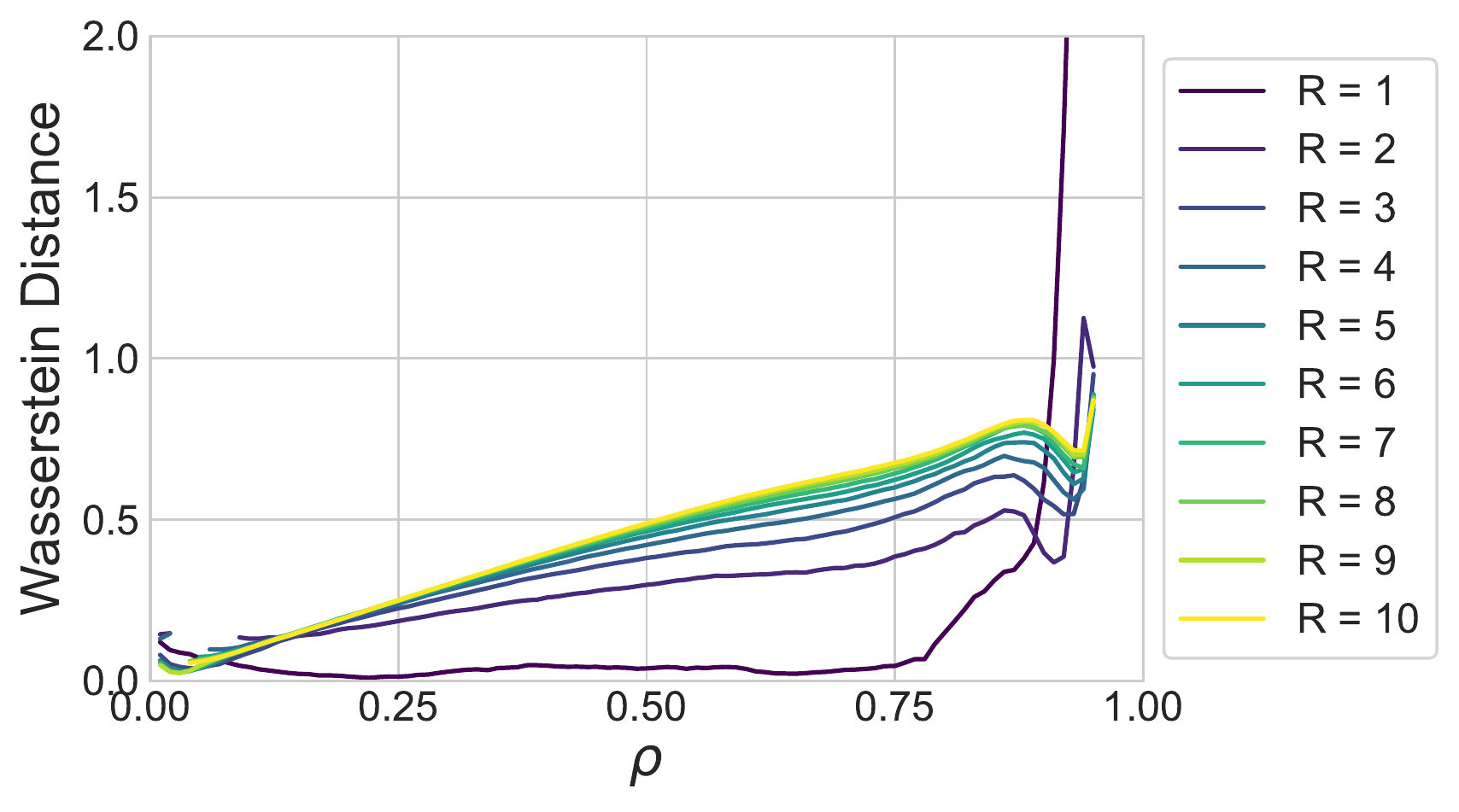}%
    \label{fig:accuracyF}}

    \caption{Accuracy of each method with 
    increasing traffic intensity $\rho$ 
    and number of CPUs $R$.}
\end{figure*}

We perform a computational experiment to compare the six
methods against one another under various circumstances.
With a fixed choice of $\mu = 1$ we calculate the sojourn
time CDFs using each method, for all $1 \leq R \leq 10$,
and all $\rho \in (0, 1)$ in steps of $0.01$.
CDFs are compared against the simulation CDF using the
Wasserstein distance \citep{mostafaei2011probability}, or
Earth-mover's distance.
This is given in \eqref{eqn:wasserstein}, with a graphic
interpretation given in Figure~\ref{fig:wasserstein}.
This measure goes from $0$, representing equal CDFs, to
$t_{\text{max}}$, the maximum sojourn time calculated,
representing the largest possible difference between the CDFs.
In practice this is calculated numerically by taking
Riemann sums with $\Delta = 0.01$ time units.

\begin{equation}\label{eqn:wasserstein}
    \Omega(G, H) = \int_{-\infty}^{+\infty} | G(t) - H(t) | dt
\end{equation}

For these experiments hyper-parameter choices are a fixed:
$L_2 = 130$; $t_{\text{max}} = 182.32$; a maximum simulation
time of $q_{\max}=160000$ time units and a warm-up time of $q_{\text{warmup}}=8000$.
The choice of the Markov chain limit $L_1$ is dependent on
$R$, it is chosen to be both large enough that the probability
of exceeding this is small, and small enough so that the
number of defined transitions is manageable, we choose
$(L_1+1)^{2R} < 10 \times 10^{10}$. For each $R$ our choice
of $L_1$ is given in Table~\ref{tbl:mc_limit}.

\begin{table}
    \centering
    \caption{Choice of Markov chain limit $L_1$ for each $R$.}
    \begin{tabular}{cccc}
      \toprule
      $R$ & $L_1$ \\
      \midrule
      1 & 22 \\
      2 & 22 \\
      3 & 22 \\
      4 & 13 \\
      5 & 7 \\
      6 & 5 \\
      7 & 4 \\
      8 & 3 \\
      9 & 3 \\
      10 & 2 \\
      \bottomrule
    \end{tabular}
    \label{tbl:mc_limit}
\end{table}


Figure.~\ref{fig:accuracyA}-\ref{fig:accuracyF} show the obtained Wasserstein distance, for each method A to F respectively, for each value of $R$ and $\rho$.

First it is important to note the scale of the y-axis on these
plots, they range from $0$ to $2$; while the Wasserstein distance
has the potential to range from $0$ to $182.32$. Therefore,
wherever the Wasserstein distance falls within the plot's range,
we can note that these are not bad approximations overall.
We can see that all methods are highly dependant on the traffic
intensity $\rho$, however the relationship between accuracy and
$\rho$ is different for the methods that use the first approximation
of $w_n$ (methods A, B and C), and those that use the second
approximation of $w_n$ (methods D, E and F). For the first
approximation, low and high values of the load $\rho$ result in
higher approximation error. This may be due to unstable
approximation algorithms used to compute matrix exponential
\citep{moler2003nineteen}. While the second approximation performs
much better for low values of $\rho$, middling values perform much
worse here. In addition, we see that the second approximation is
more dependant of $R$, with lower values of $R$ performing better.
Similarly, this dependence on $R$ is more pronounced in methods C
and F, suggesting that the second approximation of $\lambda_n$
performs worse with higher $R$ than the first Markov chain
approximation.

\begin{figure*}
    \centering
    \includegraphics[width=\textwidth]{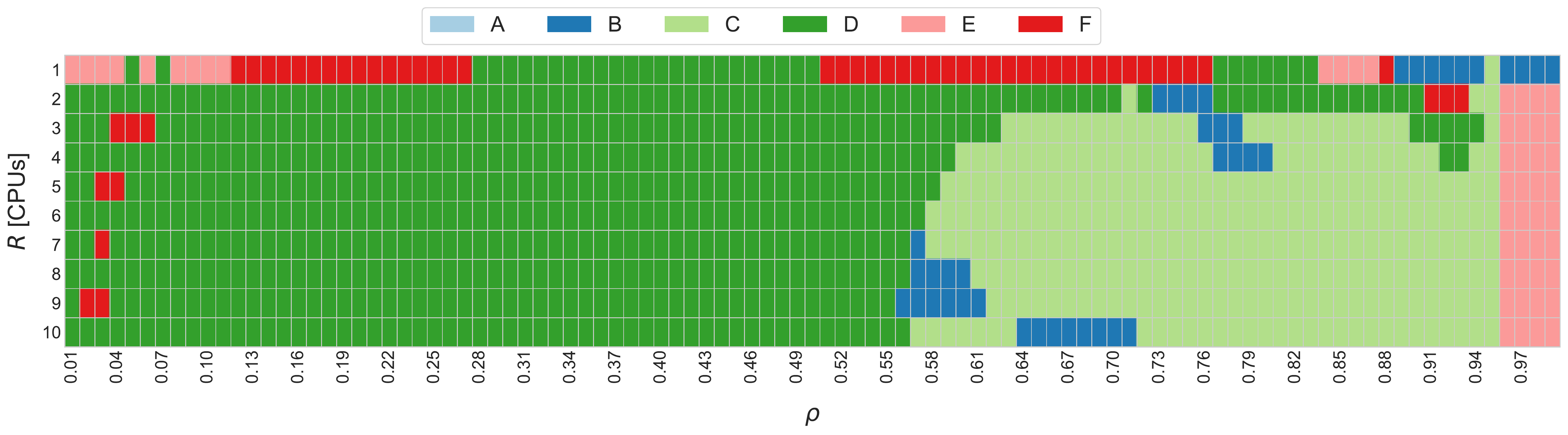}
    \caption{Most accurate method Table~\ref{tbl:methods} for each $R$, $\rho$ pair.}
    \label{fig:bestperforming}
\end{figure*}

Figure~\ref{fig:bestperforming} shows which method was most accurate
for each $R$, $\rho$ pair. From this we see that method D performed
best for low values of $\rho$,
while method C performed best
for middling to high values of $\rho$. Method E is the best performing
methods for very high values of $\rho$, however from the plot in
Figure~\ref{fig:accuracyE} we know that these are still not good
approximations of the CDF.
Interestingly, when $R=1$, that is when there is no
join-shortest-queue behaviour happening,
methods E, D and F are the best performing.

\section{Behaviour in high reliabilities}
\label{sec:behaviour-high-reliabilities}
In the prior section we have seen that methods
A-F yield an accurate approximation of the sojourn
time CDF, namely that the Wasserstein distance remains
reasonably small. Depending on the load
conditions $\rho$ we can use the approximation
with highest accuracy
(see Figure~\ref{fig:bestperforming}) to achieve
accurate sojourn time CDF approximations.

However, URLLC services ask for end-to-end latencies
with high reliabilities as
99\%, 99.9\%, 99.99\%, or 99.999\%.
This means that the network latency plus processing
latency of a service (that is the sojourn time) should be met,
e.g., 99.99\% of the times. If the end-to-end
latency requirement is of 100ms and the maximum
network latency remains below 28ms, this means that
the sojourn time should remain below 72ms in the
99.99\% of the times. Therefore, the applicability
of our methods A-F depend on their accuracy
at the 99.99\textsuperscript{th} percentile.

\begin{figure}[t]
    \begin{center}
    \includegraphics[width=\columnwidth]{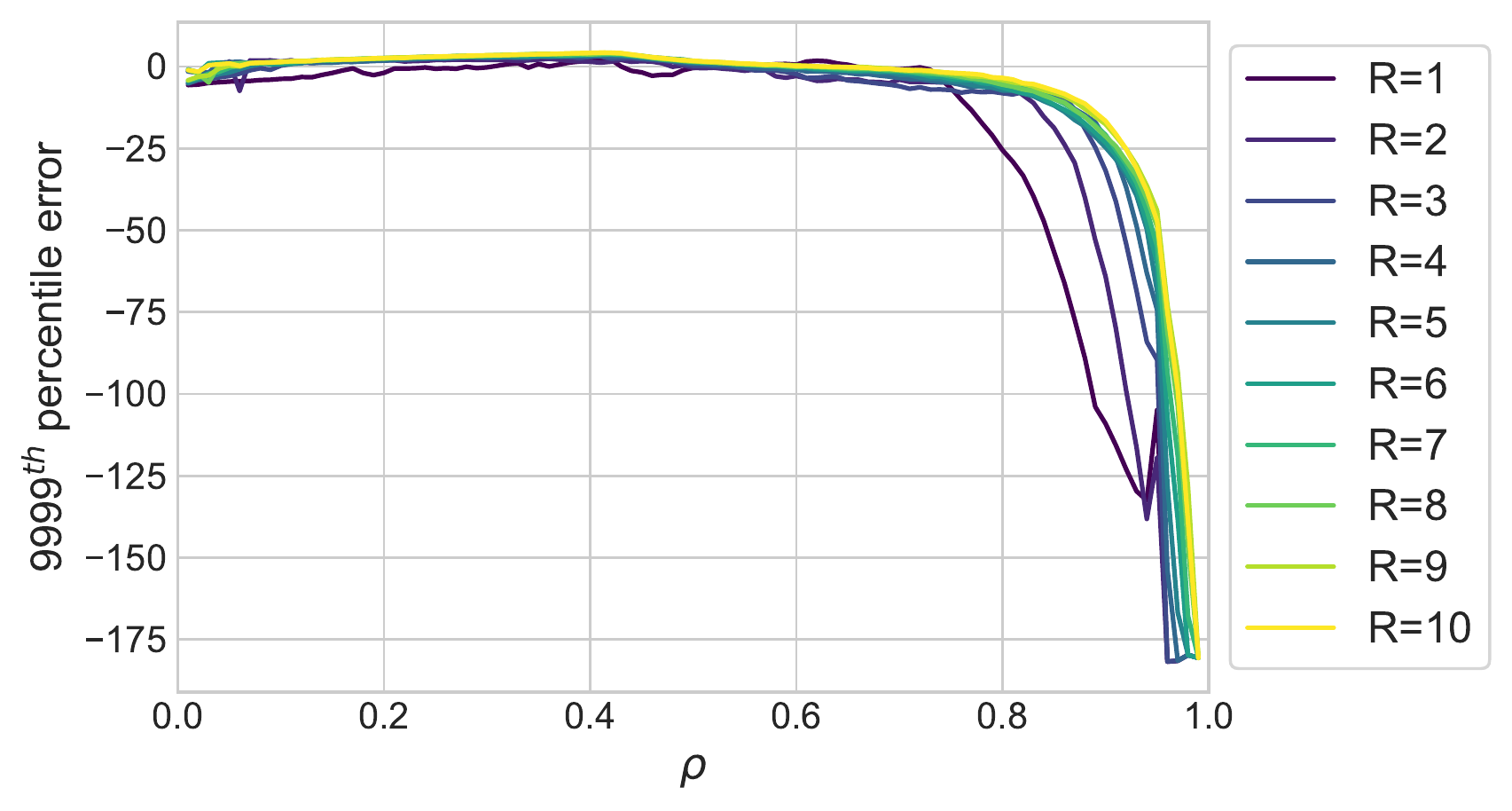}
    \end{center}
    \caption{Sojourn time 99.99\textsuperscript{th}
    percentile error using the best method.
    Positive/negative means over/under-estimation,
    respectively.}
    \label{fig:9999-percentile-sojourn-error}
\end{figure}

In Figure~\ref{fig:9999-percentile-sojourn-error}
we illustrate the error, measured in scalable
time units, achieved by the best
approximation at the 99.99\textsuperscript{th}
percentile. In other
words, if $T_{a,99.99}$ is the best method
99.99\textsuperscript{th}
percentile for the sojourn time, and
$T_{99.99}$ is the simulated 99.99-percentile;
then Figure~\ref{fig:9999-percentile-sojourn-error}
illustrates $T_{a,99.99}-T_{99.99}$. To
derive the simulated
99.99\textsuperscript{th} percentile use
to the simulation from Section~\ref{sec:simulator},

As with the Wasserstein distance
(see Figure.~\ref{fig:accuracyA}-\ref{fig:accuracyF}),
Figure~\ref{fig:9999-percentile-sojourn-error}
evidences that the 99.99\textsuperscript{th}
error becomes more prominent
as the load $\rho$ approaches to $1$ in the M/M/R-JSQ-PS system.
In particular, the best method under-estimates
the 99.99\textsuperscript{th} percentile of
the sojourn time for the error falls towards
negative values near -175 as $\rho$ approaches to $1$.
Note that the maximum sojourn time in the experiments
can pop up to $t_{\max}=182.32$ time units, hence,
the error is notoriously large towards the
highest load $\rho\approx1$.

Nevertheless, for not so high loads 
the 99.99\textsuperscript{th} percentile error
remains low. Namely, the error is of less than
$t=12$~time units with respect to the simulations
when $R\geq3$~CPUs and $\rho\leq0.85$, and
less than $t=1.78$~time units when 
$R\geq3$~CPUs and $\rho\leq0.50$.
If the system has $R<3$~CPUs, then the
best method has erratic oscillations,
indeed
the 99.99\textsubscript{th} percentile is
underestimated by $t=-56.6$~time units
with $R=1$ and $\rho=0.85$
-- see Table~\ref{tbl:percentile_errors}.

We have also analysed what is best method
error for the
99\textsuperscript{th},
99.9\textsuperscript{th}, and
99.999\textsuperscript{th} percentiles.
The results are shown in~\ref{app:percentile-errors}
and they show the same pattern as the observed
for the 99.99\textsuperscript{th} percentile
in Figure~\ref{fig:9999-percentile-sojourn-error}.
That is, the best A-F method results in
under-estimations of the sojourn time that get
worse as $\rho$ approaches to $1$. Moreover, the results from
Figure~\ref{fig:all-percentile-sojourn-errors}
in~\ref{app:percentile-errors}
shows that the error oscillations
start to become more prominent with higher
reliabilities and mid values of CPUs.

Overall, the best method
gives accurate estimations for the
99.99\textsuperscript{th} percentile of the
sojourn time as long as $\rho\leq0.85$;
tends under-estimate the 
99.99\textsuperscript{th} percentile; and
is more stable for $R\geq3$~CPUs.

\subsection{Comparison with non-exponential service times}
\label{subsec:exp-vs-others}

\begin{figure*}[t]
    \centering
    \includegraphics[width=\textwidth]{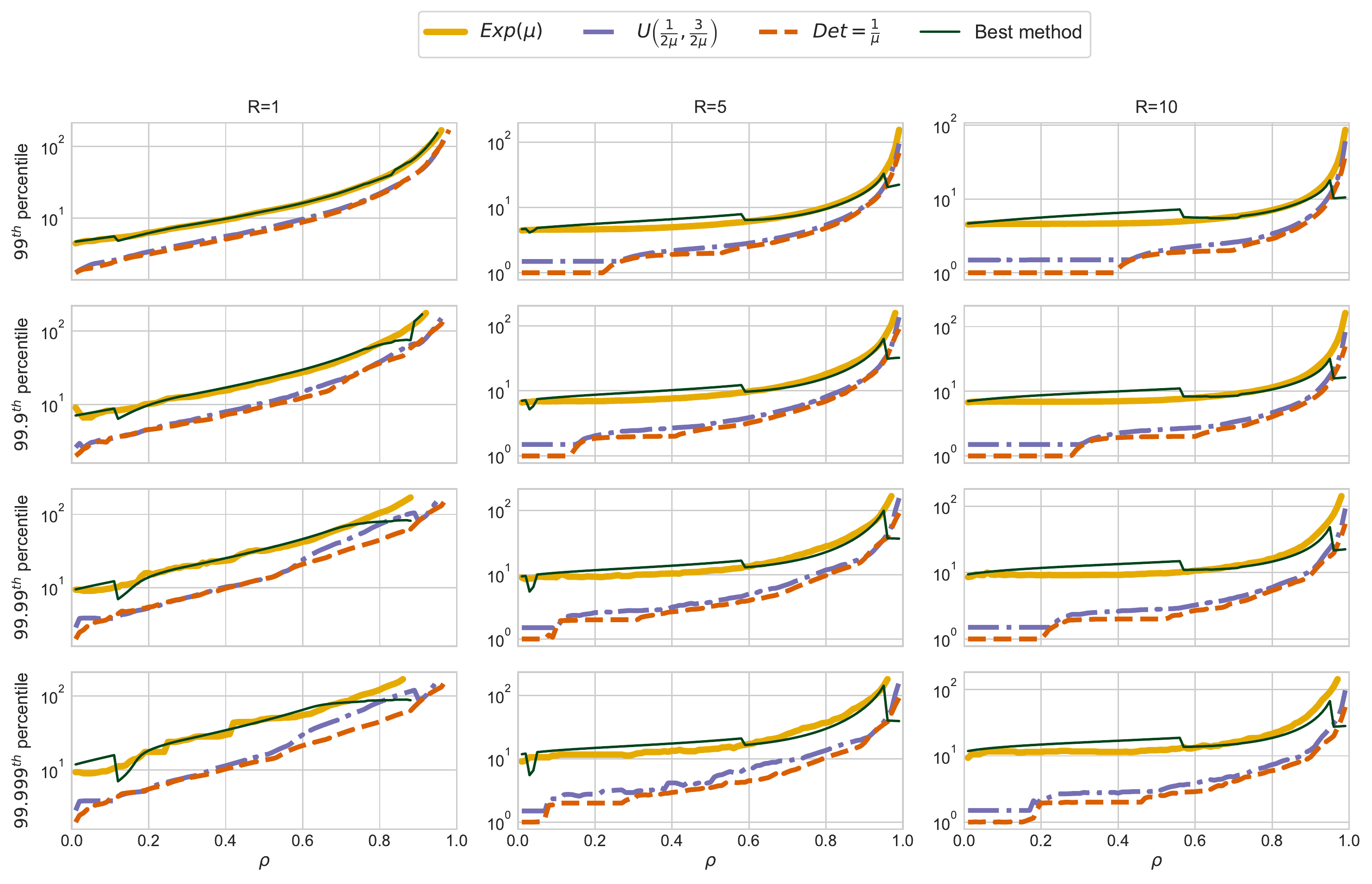}
    \caption{The 99th, 99.9th, 99.99th and 99.999th percentiles of the sojourn time distributions, on a log scale, when service times are modelled as exponentially distributed, uniformly distributed, and deterministic.}
    \label{fig:pessimism}
\end{figure*}

So far we have seen that the methods A-F
perform sufficiently well to estimate high reliabilities
of an \mbox{M/M/R-JSQ-PS} system, e.g.,
the 99.99\textsuperscript{th} percentile of the
sojourn time. However, exponentially distributed
service rates are often an unrealistic assumption,
with uniformly distributed or deterministic service
times being more realistic for services with a
bounded number of operations. Here we explore the use
of our best M/M/R-JSQ-PS approximation to upper bound
99.99\textsuperscript{th}
percentiles of the sojourn time with uniform and
deterministic service distributions. 
Such an upper bound is useful to tell whether
$R$~CPUs are enough to process URLLC service traffic,
e.g., to tell if $R$~CPUs process the URLLC service
traffic in less than 28~ms with a 99.99\% of
probability.

To investigate, we calculate the compare sojourn time
CDFs using exponentially distributed services and
uniformly distributed and deterministic services, for
various values of $R$ and $\rho$. All CDFs were
obtained using the simulation using an exponential
distribution with average service time $\tfrac{1}{\mu}$;
a uniform distribution $U\left(\tfrac{1}{2\mu},\tfrac{3}{2\mu}\right)$;
and deterministic service time of
$\tfrac{1}{\mu}$ time units. In such a manner, all
distributions share the same average service time,
although their variances are not equal: the exponentially
distributed times having the highest variance
($1/\mu^2$), followed by the uniform distributed
times ($1/(12\mu^2)$), and then the deterministic
with no variance.

We see that the CDFs obtained when modelling service times
as exponentially distributed always lie below those
obtained using uniformly distributed and deterministic
service times. This is demonstrated in Figure~\ref{fig:pessimism},
which shows that the tail percentiles are always larger, or
more pessimistic, when modelling exponential services as
opposed to uniform and deterministic services.

Figure~\ref{fig:pessimism} also evidences how
near $\rho=0.6$ when we use
$R=5$ or $R=10$~CPUs the best method (dark~green)
gets closer to the percentiles of the simulated
results (yellow). This behaviour is because
after $\rho=0.6$ the best method changes from
method~D to method~C
(see~Figure~\ref{fig:bestperforming}).
Similarly, at high loads $\rho\geq0.97$ the best
method changes from C to E, thus, the sudden
change in the sojourn time percentiles. Namely,
the sojourn time percentiles at such high loads
differs significantly from the values obtained
in the simulation (yellow). The erratic values of
the best method for $\rho\geq0.97$ even goes
below the sojourn time percentiles obtained for
uniform and deterministic service times
(blue and red lines in Figure~\ref{fig:pessimism},
respectively).

Altogether, Figure~\ref{fig:pessimism} shows: that
our best method (dark green)
stays close to the sojourn time
percentiles obtained in simulations (yellow)
for loads $\rho<0.97$;
that our best method lies above the sojourn
times provided by deterministic (red) and uniformly
distributed (blue) service times; and
that our best method largely underestimates the
sojourn time percentile for $\rho\geq0.97$,
resulting in even smaller percentiles than
uniformly distributed and deterministic service
times.



\begin{table}
\begin{center}
\caption{Sojourn time 99.99\textsuperscript{th} percentile errors using the best method with
increasing number of CPUs $R$ and load $\rho$. Positive/negative mean over/under-estimation, respectively.}
\label{tbl:percentile_errors}
\begin{tabular}{lrrrrr}
\toprule
{} &   $\pmb{R=1}$ & $\pmb{R=3}$ & $\pmb{R=5}$ & $\pmb{R=7}$ & $\pmb{R=10}$\\
\midrule
$\pmb{\rho = 0.10}$ & -3.76 & 0.29 & 1.49 & 1.38 & 1.50\\
$\pmb{\rho = 0.25}$ & -0.68 & 2.38 & 2.58 & 2.94 & 3.04\\
$\pmb{\rho = 0.50}$ & -0.41 & 0.22 & 1.17 & 1.47 & 1.78\\
$\pmb{\rho = 0.75}$ & -6.56 & -7.15 & -4.51 & -3.04 & -1.87\\
$\pmb{\rho=0.85}$ & -56.60 &  -8.83 & -11.87 & -9.84 & -7.29\\
$\pmb{\rho = 0.90}$ & -109.06 & -31.78 & -24.69 & -20.86 & -16.63\\
$\pmb{\rho = 0.95}$ & -104.87 & -89.57 & -68.65 & -54.94 & -47.24\\
$\pmb{\rho = 0.99}$ & -180.58 & -180.59 & -180.60 & -180.61 & -180.62\\

\bottomrule
\end{tabular}
\end{center}
\end{table}

\section{Conclusions}
\label{sec:conclusions}

This paper models the M/M/R-JSQ-PS sojourn time
distribution for URLLC services whose traffic
is processed using a multi-processor PS system.
In the paper we:
\begin{enumerate}[i)]
    \item present a generic open-source discrete event
    simulation software for \mbox{G/G/R-JSQ-PS} systems;
    \item derive and compare six analytical approximations
    for the sojourn time CDF of \mbox{M/M/R-JSQ-PS} systems,
    and analyse their run time complexities; and
    \item investigate the applicability of \mbox{M/M/R-JSQ-PS}
    models to \mbox{M/G/R-JSQ-PS} systems under both Uniform
    and Deterministic intended service times.
\end{enumerate}

The proposed approximations have
polynomial time complexities $\mathcal{O}\left(L_1^{3R}\right)$,
and are useful to determine if $R$~CPUs
are enough to meet the URLLC requirements
under mid loads, for they yield errors of 
less than 1.78~time units in high
percentiles as a 99.99\%. For mid to high loads
the error remains below 12~time units.

\section*{Acknowledgements}
This work has been partially funded by European Union’s
Horizon 2020 research and innovation programme under grant agreement
No 101015956, and the Spanish Ministry of Economic Affairs and Digital
Transformation and the European Union-NextGenerationEU through the
UNICO 5G I+D 6G-EDGEDT and 6G-DATADRIVEN. 




\appendix

\renewcommand*{\thesection}{Appendix~\Alph{section}}

\section{Errors for increasing
reliabilities}
\label{app:percentile-errors}

\begin{figure}[ht!]
    \centering

    \subfloat[]{\includegraphics[width=0.4\textwidth]{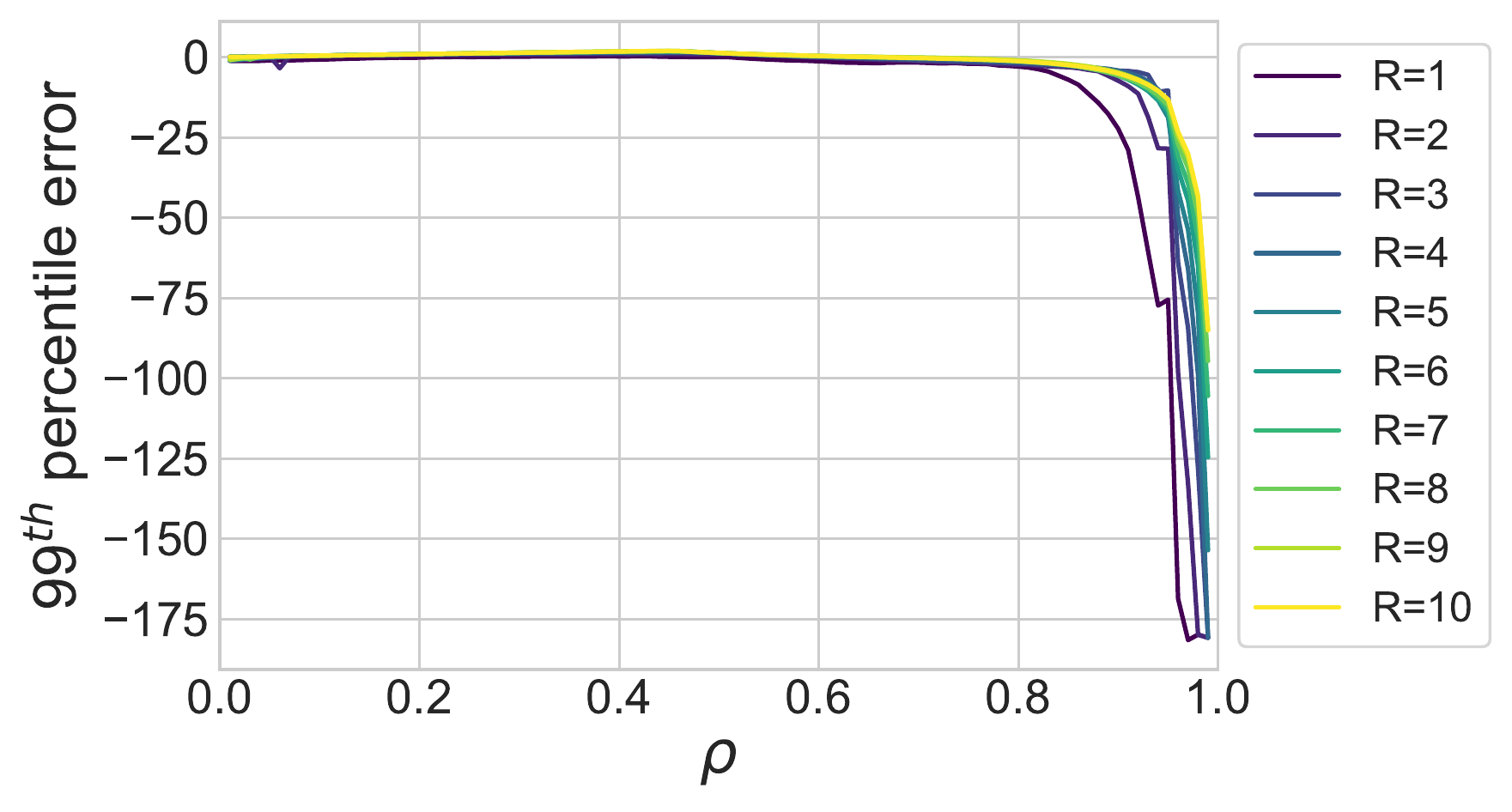}}\\
    \subfloat[]{\includegraphics[width=0.4\textwidth]{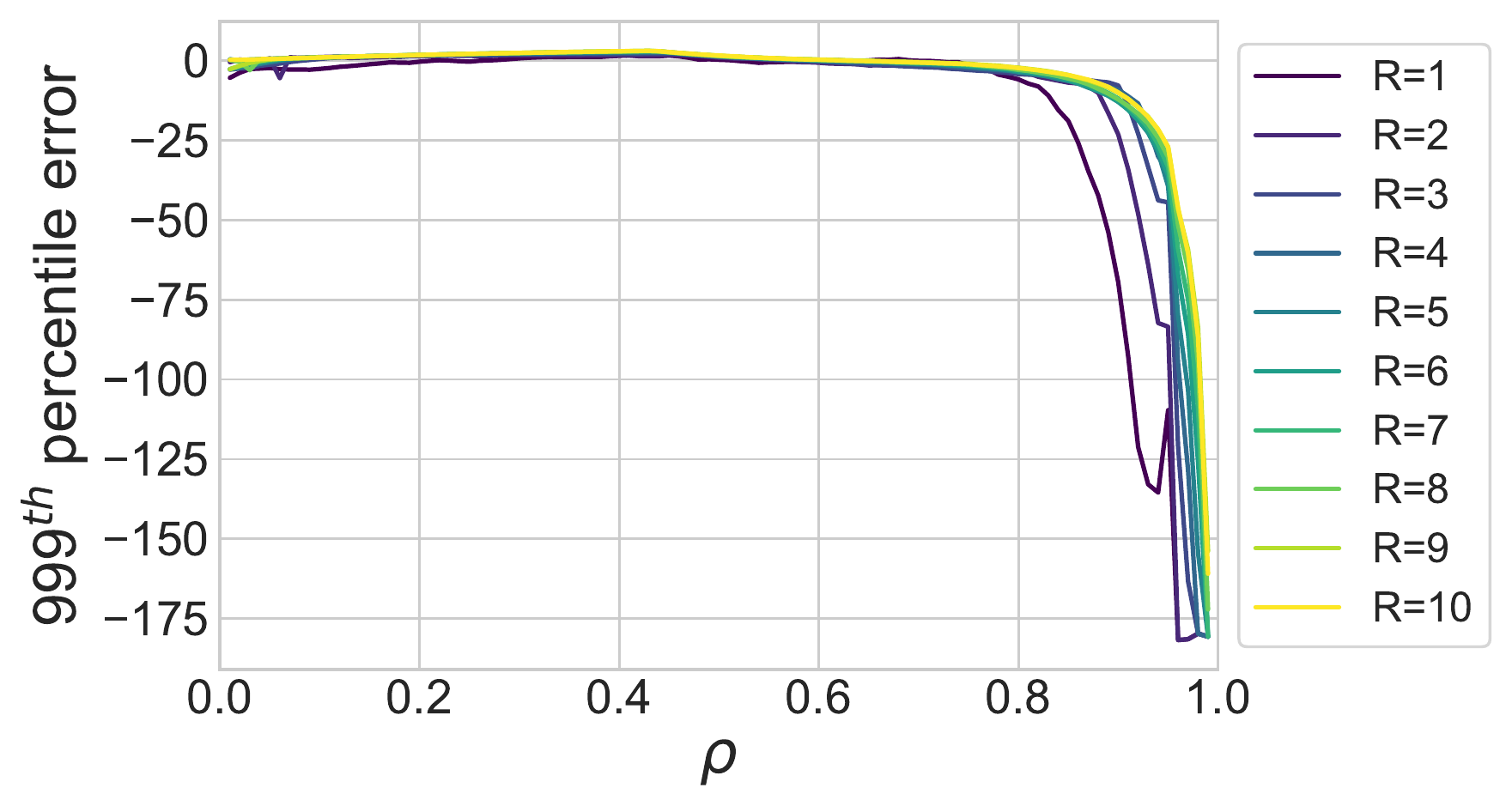}} \\
    \subfloat[]{\includegraphics[width=0.4\textwidth]{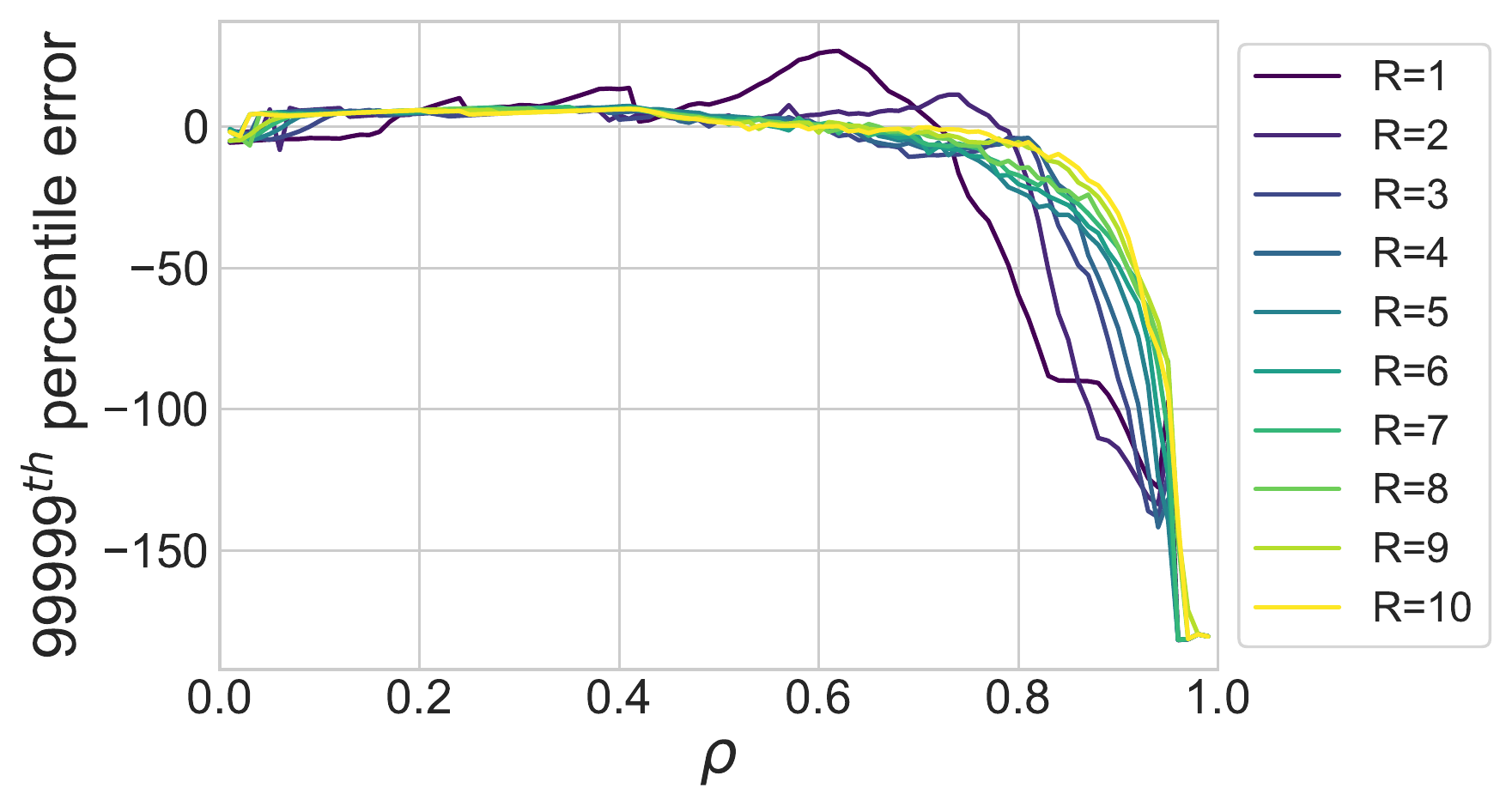}} \\

    \caption{The best method error for the
        99\textsuperscript{th},
        99.9\textsuperscript{th}, and
        99.999\textsuperscript{th}
        percentile the sojourn time.
 Positive/negative mean over/under-estimation, respectively.
    }
    \label{fig:all-percentile-sojourn-errors}
\end{figure}

Figure~\ref{fig:all-percentile-sojourn-errors}
shows the error for the best
approximation A-F
from the 99\textsuperscript{th} up to
the 99.999\textsuperscript{th} percentile of the
sojourn time. The sojourn time error is unitless,
and it illustrates the increasing error as
$\rho$ approaches $1$, so as the increasing oscillations of
the error for higher reliabilities, even with
mid values of the number of CPUs like $R=4$ --
as explained in
Section~\ref{sec:behaviour-high-reliabilities}.

\section{Getting the sojourn time percentile}
\label{app:sojourn-percentile}

We provide an open-source implementation\footnote{\url{https://github.com/geraintpalmer/mmr-jsq-ps/}}
of the proposed approximation methods~A-F.
Every method is implemented in Python and
yields the sojourn time CDF for a given
number of CPUs $R$. Additionally, it is possible
to specify the truncation limits for the maximum number
of customers considered at each CPU $L_1$, and the
maximum number of customers at the system $L_2$.

In order to obtain the
sojourn time $\eta$-percentile with $R$ CPUs,
we first compute the load $\rho$ given
the $R$ CPUs, and the arrival and
service rates $\Lambda,\mu$; respectively.
Second, we check
Figure~\ref{fig:bestperforming}
to know 
which is the best method for the given $(\rho,R)$ tuple,
e.g., method-A.
Third, we create an instance of method-A invoking
\begin{lstlisting}
jsq.MethodA($\Lambda,\mu,R,L_1,L_2, \{t_0,t_1,\ldots\}$)
\end{lstlisting}
with $\{t_0,t_1,\ldots\}$ being the discrete time points
at which we compute the CDF.
Then, we obtain the CDF of method-A
by accessing property \texttt{sojourn\_time\_cdf} of the method instance.
This property holds a vector
$\{P_0,P_1,\ldots\}$ that represents the
CDF computed by method-A. In particular, each element
represents $P_i=\mathbb{P}(T\leq t_i)$.
Finally, we obtain the $\eta$-percentile of the sojourn time
(denoted as $t^\eta$) as
\begin{equation}
    t^{\eta}=\argmin\{t_i: \mathbb{P}(T\leq t_i)>\eta\}
\end{equation}

\printcredits

\bibliographystyle{cas-model2-names}

\bibliography{bibliography.bib}

\balance

\bio{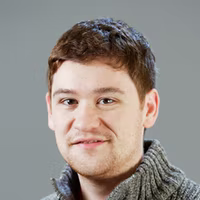}
Geraint I. Palmer
graduated from Aberystwyth University in 2013 with a BSc in mathematics, and then moved to
Cardiff University to obtain his MSc in
operational research and applied statistics in 2014, and
his PhD in applied stochastic modelling in 2018, for which he won the OR Society's Doctoral Award. He now works as a lecturer
at Cardiff University where his research is in
operational research, queueing models and discrete
event simulation.
\endbio

\bio{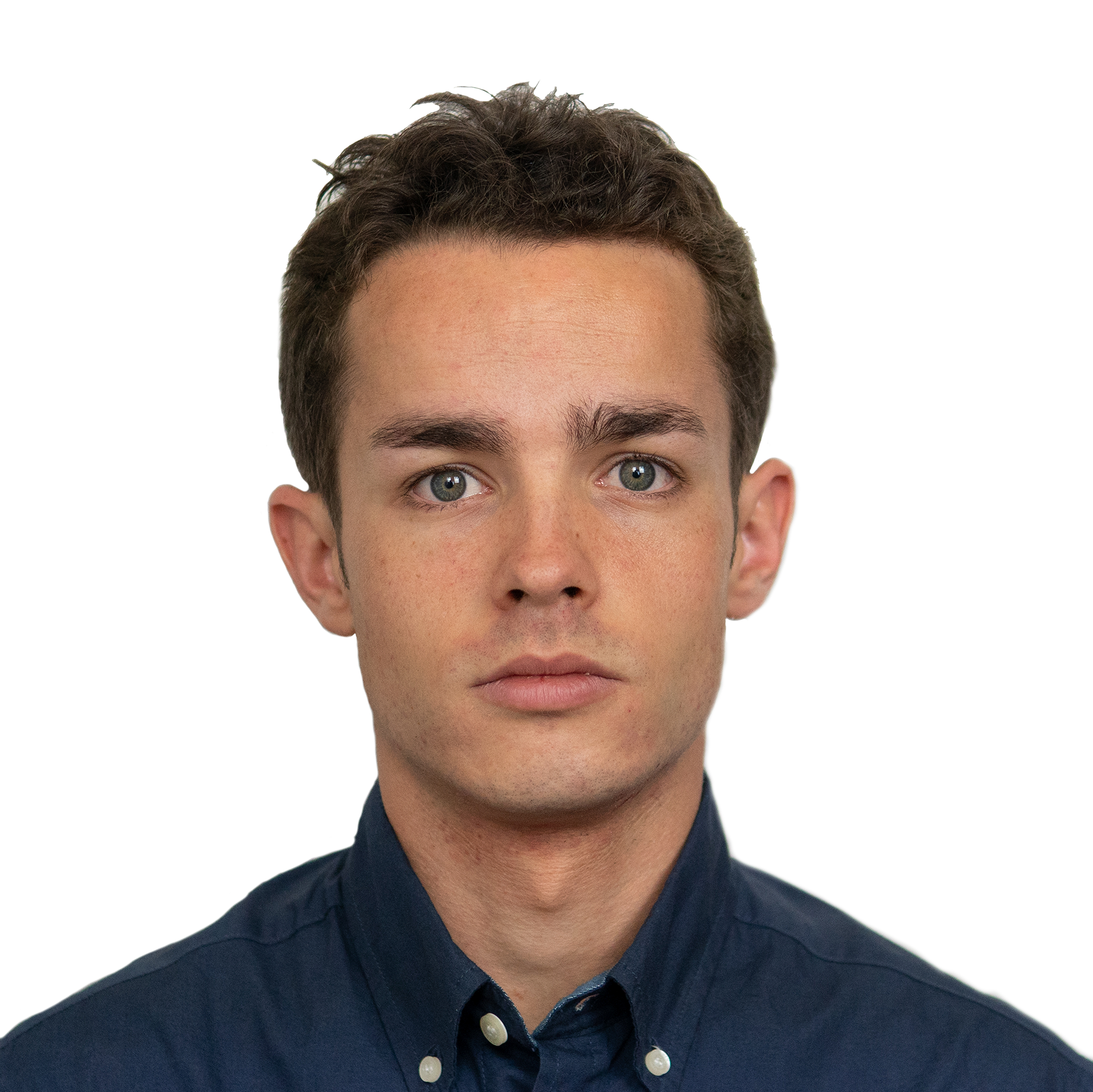}
Jorge Mart\'in P\'erez obtained a B.Sc in mathematics, and
a B.Sc in computer science, both at Universidad
Autónoma de Madrid (UAM) in 2016. He obtained his M.Sc. and Ph.D in
Telematics from Universidad Carlos III de Madrid (UC3M) in 2017 and 2021, respectively.
His research focuses in optimal resource allocation in
networks, and since 2016 he participates in
EU funded research projects in UC3M Telematics department.
\endbio

\end{document}

%% file: mmc.tex
\begin{tikzpicture}[>=latex,draw=black,text=black]

\tikzset{
myshape/.style={
  rectangle split,
  minimum height=2cm,
  minimum width=1.5cm,
  rectangle split parts=3,
  draw,
  anchor=center,
  },
packet/.style={
  draw,
  anchor=center,
  rectangle,
  draw,
  minimum width=.5cm,
  minimum height=.2cm,
  fill=gray!40
  },
mytri/.style={
  draw,
  shape=rectangle,
  minimum height=1.5cm,
  minimum width=1.5cm,
  }
}


\node[packet,label=above:{URLLC packets}] at (1.4,0.3) {};
\node[packet] at (2.1,0.3) {};
\node[packet] at (0.7,0.3) {};

\draw [->]  (0,0) -- node[label=below:{$\Lambda$ Mbps}] {} (3,0);
\node[align=center, rectangle, draw] at (4,0) {JSQ\\ scheduler};

\draw[->] (5,0) -- (5.5,1.5) -- node[label=below:{$\lambda_1$ Mbps}] {} (7.3,1.5);

\node[packet] at (8,1.1) {};
\node[packet] at (8,1.5) {};
\draw (7.5,2.2) -- (8.5,2.2) --  (8.5,.8) -- (7.5,.8);
\node[draw,circle,inner sep=.3cm] at (9.3,1.5) {$\mu$};
\draw[->] (10,1.5) -- (10.5,1.5);
\node at (8.8,.5) {PS server 1};

\draw (5,0) -- (6.1,0);
\node at (6.7,0) {$\ldots$};
\node[rotate=90] at (8.8,-.2) {$\ldots$};

\draw[->] (5,0) -- (5.5,-1.5) -- node[label=below:{$\lambda_R$ Mbps}] {} (7.3,-1.5);

\node[packet] at (8,-1.9) {};
\draw (7.5,-2.2) -- (8.5,-2.2) --  (8.5,-.8) -- (7.5,-.8);
\node[draw,circle,inner sep=.3cm] at (9.3,-1.5) {$\mu$};
\draw[->] (10,-1.5) -- (10.5,-1.5);
\node at (8.8,-2.5) {PS server R};

\end{tikzpicture}

%% file: img/eventschedulingapproach.tex
\begin{tikzpicture}

\node[align=center] (start) [style={minimum size=2cm, text width=1.8cm, draw=black,fill=black!10,text=black,shape=circle}] at (0, 19) {Start};
\node[align=center] (initial) [style={minimum width=5cm, minimum height=1cm, text width=4.8cm, draw=black,fill=black!10,text=black,shape=rectangle}] at (0, 16.5) {\small{Initialise Simulation}};
\node[align=center] (Aphase) [style={minimum width=5cm, minimum height=1cm, text width=4.8cm, draw=black,fill=black!10,text=black,shape=rectangle}] at (0, 14.5) {\small{\textbf{A}-Phase}};
\node[align=center] (Bphase) [style={minimum width=5cm, minimum height=1cm, text width=4.8cm, draw=black,fill=black!10,text=black,shape=rectangle}] at (0, 12.5) {\small{\textbf{B}-Phase}};
\node[align=center] (Cphase) [style={minimum width=5cm, minimum height=1cm, text width=4.8cm, draw=black,fill=black!10,text=black,shape=rectangle}] at (0, 10.5) {\small{\textbf{C}-Phase}};
\node[align=center] (anymoreC) [style={minimum size=2cm, text width=1.8cm, draw=black,fill=black!10,text=black,shape=diamond, aspect=2}] at (0, 8) {\small{Any more \textbf{C}-Events?}};
\node[align=center] (anymoreB) [style={minimum size=2cm, text width=1.8cm, draw=black,fill=black!10,text=black,shape=diamond, aspect=2}] at (0, 5) {\small{Any more \textbf{B}-Events?}};
\node[align=center] (isend) [style={minimum size=2cm, text width=1.8cm, draw=black,fill=black!10,text=black,shape=diamond, aspect=2}] at (0, 2) {\small{Simulation Complete?}};
\node[align=center] (stop) [style={minimum size=2cm, text width=1.8cm, draw=black,fill=black!10,text=black,shape=circle}] at (0, -1) {\small{Stop}};

\draw[ultra thick, -triangle 90] (start) -- (initial);
\draw[ultra thick, -triangle 90] (initial) -- (Aphase);
\draw[ultra thick, -triangle 90] (Aphase) -- (Bphase);
\draw[ultra thick, -triangle 90] (Bphase) -- (Cphase);
\draw[ultra thick, -triangle 90] (Cphase) -- (anymoreC);
\draw[ultra thick, -triangle 90] (anymoreC) -- (anymoreB);
\draw[ultra thick, -triangle 90] (anymoreB) -- (isend);
\draw[ultra thick, -triangle 90] (isend) -- (stop);

\draw[ultra thick, -triangle 90] (anymoreC) -- (-4, 8) -- (-4, 10.5) -- (Cphase);
\draw[ultra thick, -triangle 90] (anymoreB) -- (-4.75, 5) -- (-4.75, 12.5) -- (Bphase);
\draw[ultra thick, -triangle 90] (isend) -- (-5.5, 2) -- (-5.5, 14.5) -- (Aphase);

\node at (0.5, 6.8) {No};
\node at (-5, 2.5) {No};
\node at (0.5, 3.8) {No};
\node at (-3.5, 8.5) {Yes};
\node at (-4.25, 5.5) {Yes};
\node at (0.5, 0.8) {Yes};

\end{tikzpicture}

%% file: markov_chain.tex
\begin{standalone}
\begin{tikzpicture}[draw=black,text=black]
    \tikzstyle{state}=[minimum width=1.5cm, font=\boldmath];
    \node (00) at (0,0) [state] {$(0, 0)$};
    \node (01) at ($(00)+(3,0)$) [state] {$(0, 1)$};
    \node (02) at ($(01)+(3,0)$) [state] {$(0, 2)$};
    \node (03) at ($(02)+(3,0)$) [state] {$(0, 3)$};
    \node (04h) at ($(03)+(2.25,0)$) [state] {};

    \node (10) at ($(00)+(0,-3)$) [state] {$(1, 0)$};
    \node (11) at ($(10)+(3,0)$) [state] {$(1, 1)$};
    \node (12) at ($(11)+(3,0)$) [state] {$(1, 2)$};
    \node (13) at ($(12)+(3,0)$) [state] {$(1, 3)$};
    \node (14h) at ($(13)+(2.25,0)$) [state] {};

    \node (20) at ($(10)+(0,-3)$) [state] {$(2, 0)$};
    \node (21) at ($(20)+(3,0)$) [state] {$(2, 1)$};
    \node (22) at ($(21)+(3,0)$) [state] {$(2, 2)$};
    \node (23) at ($(22)+(3,0)$) [state] {$(2, 3)$};
    \node (24h) at ($(23)+(2.25,0)$) [state] {};

    \node (30) at ($(20)+(0,-3)$) [state] {$(3, 0)$};
    \node (31) at ($(30)+(3,0)$) [state] {$(3, 1)$};
    \node (32) at ($(31)+(3,0)$) [state] {$(3, 2)$};
    \node (33) at ($(32)+(3,0)$) [state] {$(3, 3)$};
    \node (34h) at ($(33)+(2.25,0)$) [state] {};
    \node (34q) at ($(34h)+(-0.15,+0.6)$) [state] {};

    \node (40h) at ($(30)+(0,-1.25)$) [state] {};
    \node (41h) at ($(40h)+(3,0)$) [state] {};
    \node (42h) at ($(41h)+(3,0)$) [state] {};
    \node (43h) at ($(42h)+(3,0)$) [state] {};
    \node (43q) at ($(43h)+(-0.6,0)$) [state] {};

    \draw (00) edge[out=45,in=135,->,thick] node [below] {\footnotesize$\frac{1}{2}\Lambda$} (01);
    \draw (10) edge[out=45,in=135,->,thick] node [below] {\footnotesize$\Lambda$} (11);
    \draw (11) edge[out=45,in=135,->,thick] node [below] {\footnotesize$\frac{1}{2}\Lambda$} (12);
    \draw (20) edge[out=45,in=135,->,thick] node [below] {\footnotesize$\Lambda$} (21);
    \draw (21) edge[out=45,in=135,->,thick] node [below] {\footnotesize$\Lambda$} (22);
    \draw (22) edge[out=45,in=135,->,thick] node [below] {\footnotesize$\frac{1}{2}\Lambda$} (23);
    \draw (30) edge[out=45,in=135,->,thick] node [below] {\footnotesize$\Lambda$} (31);
    \draw (31) edge[out=45,in=135,->,thick] node [below] {\footnotesize$\Lambda$} (32);
    \draw (32) edge[out=45,in=135,->,thick] node [below] {\footnotesize$\Lambda$} (33);
    \draw (33) edge[out=45,in=180,->,thick] node [above] {\footnotesize$\frac{1}{2}\Lambda$} (34q);

    \draw (00) edge[out=-135,in=135,->,thick] node [right] {\footnotesize$\frac{1}{2}\Lambda$} (10);
    \draw (01) edge[out=-135,in=135,->,thick] node [right] {\footnotesize$\Lambda$} (11);
    \draw (11) edge[out=-135,in=135,->,thick] node [right] {\footnotesize$\frac{1}{2}\Lambda$} (21);
    \draw (02) edge[out=-135,in=135,->,thick] node [right] {\footnotesize$\Lambda$} (12);
    \draw (12) edge[out=-135,in=135,->,thick] node [right] {\footnotesize$\Lambda$} (22);
    \draw (22) edge[out=-135,in=135,->,thick] node [right] {\footnotesize$\frac{1}{2}\Lambda$} (32);
    \draw (03) edge[out=-135,in=135,->,thick] node [right] {\footnotesize$\Lambda$} (13);
    \draw (13) edge[out=-135,in=135,->,thick] node [right] {\footnotesize$\Lambda$} (23);
    \draw (23) edge[out=-135,in=135,->,thick] node [right] {\footnotesize$\Lambda$} (33);
    \draw (33) edge[out=-135,in=90,->,thick] node [left] {\footnotesize$\frac{1}{2}\Lambda$} (43q);

    \draw (01) edge[->,thick] node [below] {\footnotesize$\mu$} (00);
    \draw (02) edge[->,thick] node [below] {\footnotesize$\mu$} (01);
    \draw (03) edge[->,thick] node [below] {\footnotesize$\mu$} (02);
    \draw (11) edge[->,thick] node [below] {\footnotesize$\mu$} (10);
    \draw (12) edge[->,thick] node [below] {\footnotesize$\mu$} (11);
    \draw (13) edge[->,thick] node [below] {\footnotesize$\mu$} (12);
    \draw (21) edge[->,thick] node [below] {\footnotesize$\mu$} (20);
    \draw (22) edge[->,thick] node [below] {\footnotesize$\mu$} (21);
    \draw (23) edge[->,thick] node [below] {\footnotesize$\mu$} (22);
    \draw (31) edge[->,thick] node [below] {\footnotesize$\mu$} (30);
    \draw (32) edge[->,thick] node [below] {\footnotesize$\mu$} (31);
    \draw (33) edge[->,thick] node [below] {\footnotesize$\mu$} (32);
    
    \draw (04h) edge[->,thick] node [below] {\footnotesize$\mu$} (03);
    \draw (14h) edge[->,thick] node [below] {\footnotesize$\mu$} (13);
    \draw (24h) edge[->,thick] node [below] {\footnotesize$\mu$} (23);
    \draw (34h) edge[->,thick] node [below] {\footnotesize$\mu$} (33);

    \draw (10) edge[->,thick] node [right] {\footnotesize$\mu$} (00);
    \draw (20) edge[->,thick] node [right] {\footnotesize$\mu$} (10);
    \draw (30) edge[->,thick] node [right] {\footnotesize$\mu$} (20);
    \draw (11) edge[->,thick] node [right] {\footnotesize$\mu$} (01);
    \draw (21) edge[->,thick] node [right] {\footnotesize$\mu$} (11);
    \draw (31) edge[->,thick] node [right] {\footnotesize$\mu$} (21);
    \draw (12) edge[->,thick] node [right] {\footnotesize$\mu$} (02);
    \draw (22) edge[->,thick] node [right] {\footnotesize$\mu$} (12);
    \draw (32) edge[->,thick] node [right] {\footnotesize$\mu$} (22);
    \draw (13) edge[->,thick] node [right] {\footnotesize$\mu$} (03);
    \draw (23) edge[->,thick] node [right] {\footnotesize$\mu$} (13);
    \draw (33) edge[->,thick] node [right] {\footnotesize$\mu$} (23);

    \draw (40h) edge[->,thick] node [right] {\footnotesize$\mu$} (30);
    \draw (41h) edge[->,thick] node [right] {\footnotesize$\mu$} (31);
    \draw (42h) edge[->,thick] node [right] {\footnotesize$\mu$} (32);
    \draw (43h) edge[->,thick] node [right] {\footnotesize$\mu$} (33);

\end{tikzpicture}
\end{standalone}

%% file: wasserstein.tex
\begin{tikzpicture}[draw=black,text=black]
\begin{axis}[axis x line=bottom,axis y line=left,xlabel={Sojourn time}, ylabel={CDF},
    legend cell align={left},
    legend pos=south east,
    xmin=0, xmax=6.6,
    ymin=0, ymax=1.1,
    xtick={0,  6.4},
    xticklabels={$0$, $t_{\text{max}}$},
    ytick={0, 1},
]
 
\addplot [name path = A,
    thick,
    domain = 0:6,
    samples = 1000] {1-exp(-2*x)};
 \addlegendentry{Actual}

\addplot [name path = B,
    thick, dashed,
    domain = 0:6,
    samples = 1000] {1-exp(-x)};
 \addlegendentry{Approximation}

\addplot [black!20, area legend] fill between [of = A and B];
\addlegendentry{Wasserstein distance}


\end{axis}
\end{tikzpicture}